 \documentclass[10pt,preprint]{aastex}

\newcommand{\Ell}{E_\parallel}      
\newcommand{\rhoGJ}{\rho_{{\rm GJ}}}  
\newcommand{\sgT}{\sigma_{\rm T}}  
\newcommand{\sgP}{\sigma_{\rm p}}  
\newcommand{\rlc}{\varpi_{\rm LC}} 
\newcommand{\inc}{\alpha_{\rm i}}  
\newcommand{\etaICe}{\eta_{\rm IC}^{\rm e}}
\newcommand{\etaICg}{\eta_{\rm IC}^\gamma}
\newcommand{\etaSC}{\eta_{\rm SC}}
\newcommand{\etaP}{\eta_{\rm p}}
\newcommand{\bCM}{\beta_{\rm CM}} %
\newcommand{\rhowSQR}{\rho_{\rm w}^2}

 \slugcomment{Submitted to Astroph. J. on April 3, 2006}

\shorttitle{Particle Accelerator in Pulsar Magnetospheres}
\shortauthors{Hirotani}

\begin{document}


\title{Particle Accelerator in Pulsar Magnetospheres:
       Super Goldreich-Julian Current with 
       Ion Emission from the Neutron Star Surface}


\author{Kouichi Hirotani}
\affil{ASIAA/National Tsing Hua University - TIARA,\\
       PO Box 23-141, Taipei, Taiwan\footnote{
            Postal address: 
            TIARA, Department of Physics, 
            National Tsing Hua University,
            101, Sec. 2, Kuang Fu Rd.,Hsinchu, Taiwan 300}
      }
\email{hirotani@tiara.sinica.edu.tw}




\begin{abstract}
We investigate the self-consistent electrodynamic structure of 
a particle accelerator in the Crab pulsar magnetosphere
on the two-dimensional poloidal plane,
solving the Poisson equation for the electrostatic potential
together with the Boltzmann equations for 
electrons, positrons and gamma-rays.
If the trans-field thickness of the gap is thin,
the created current density becomes sub-Goldreich-Julian,
giving the traditional outer-gap solution
but with negligible gamma-ray luminosity.
As the thickness increases, the created current increases to become
super-Goldreich-Julian, 
giving a new gap solution with substantially screened
acceleration electric field in the inner part.
In this case, the gap extends towards the neutron star
with a small-amplitude positive acceleration field,
extracting ions from the stellar surface 
as a space-charge-limited flow.
The acceleration field is highly unscreened in the outer magnetosphere,
resulting in a gamma-ray spectral shape which is consistent with 
the observations.
\end{abstract}





\section{Introduction}
\label{sec:intro}
%
The Energetic Gamma Ray Experiment Telescope (EGRET) 
aboard the Compton Gamma Ray Observatory 
has detected pulsed signals from at least six rotation-powered pulsars
(e.g., for the Crab pulsar, Nolan et al. 1993, Fierro et al. 1998).
Since interpreting $\gamma$-rays should be less ambiguous
compared with reprocessed, non-thermal X-rays,
the $\gamma$-ray pulsations observed from these objects
are particularly important as a direct signature of 
basic non-thermal processes in pulsar magnetospheres,
and potentially should help to discriminate 
among different emission models.

\begin{figure}
 \epsscale{0.6}
 \plotone{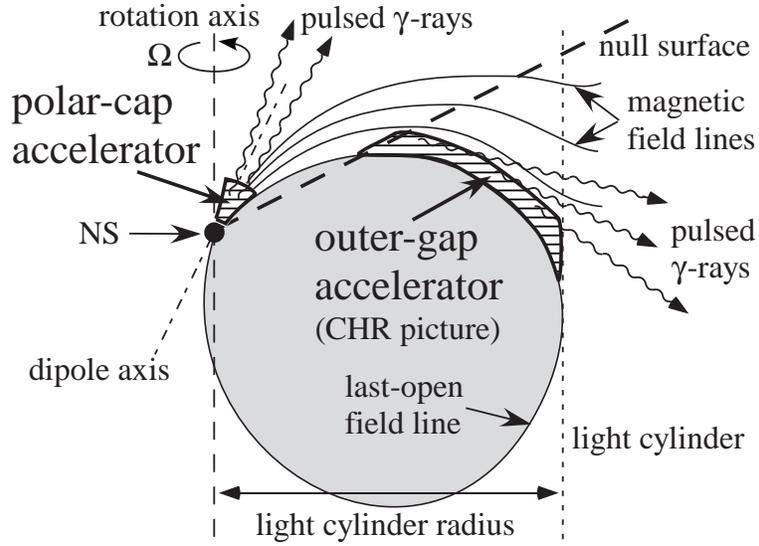}
\caption{
Schematic figure (sideview) of 
the two representative accelerator models. 
The small filled circle represents the neutron star
\label{fig:sidev}
}
\end{figure}

The pulsar magnetosphere can be divided into two zones 
(fig.~\ref{fig:sidev}):
The closed zone filled with a dense plasma corotating with the star,
and the open zone in which plasma flows along the open field lines
to escape through the light cylinder.
The last-open field lines form the border of the open magnetic field
line bundle.
In all the pulsar emission models, 
particle acceleration and the resultant photon emissions
take place within this open zone.

\begin{figure}
\epsscale{0.8}
\plottwo{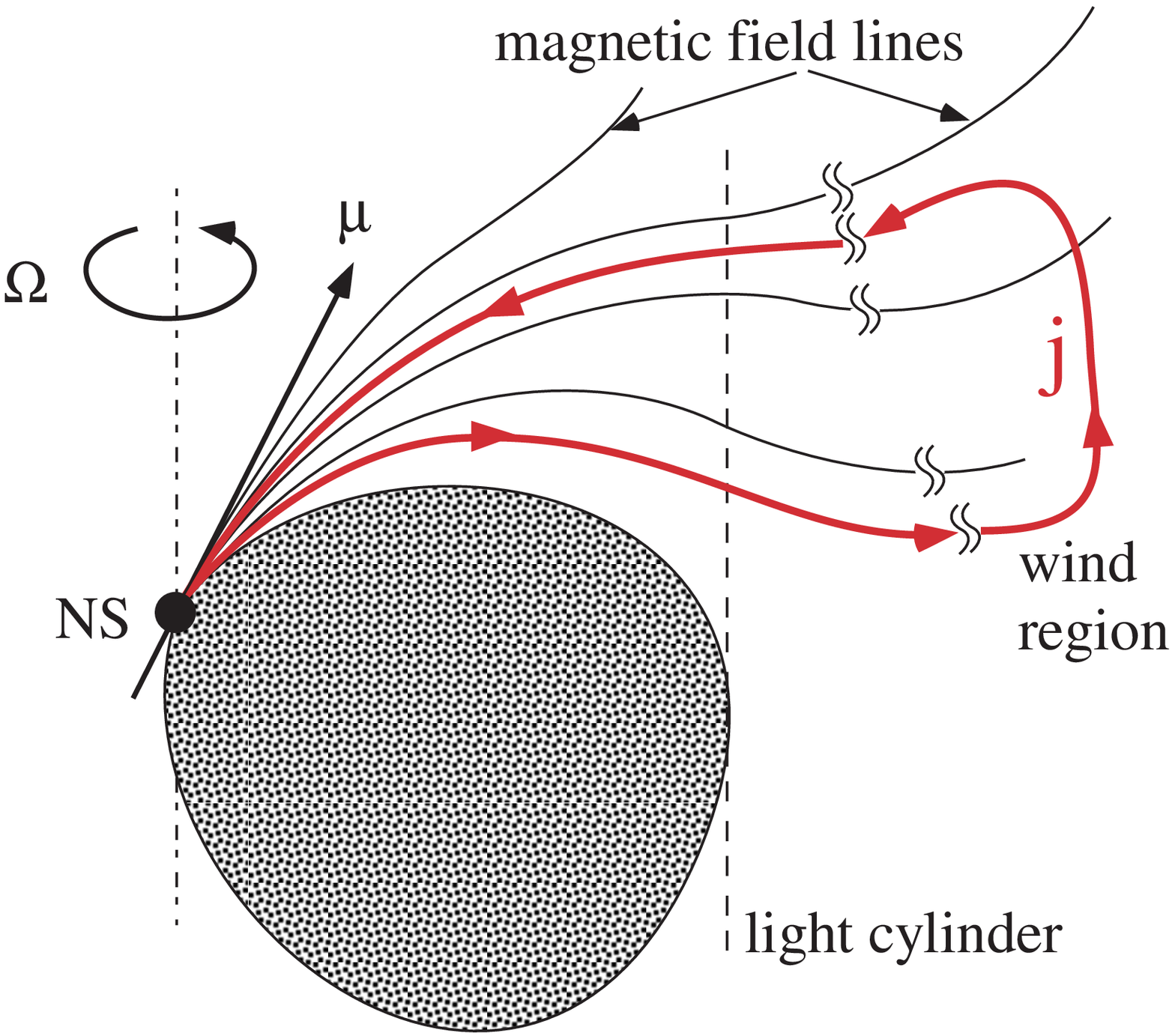}{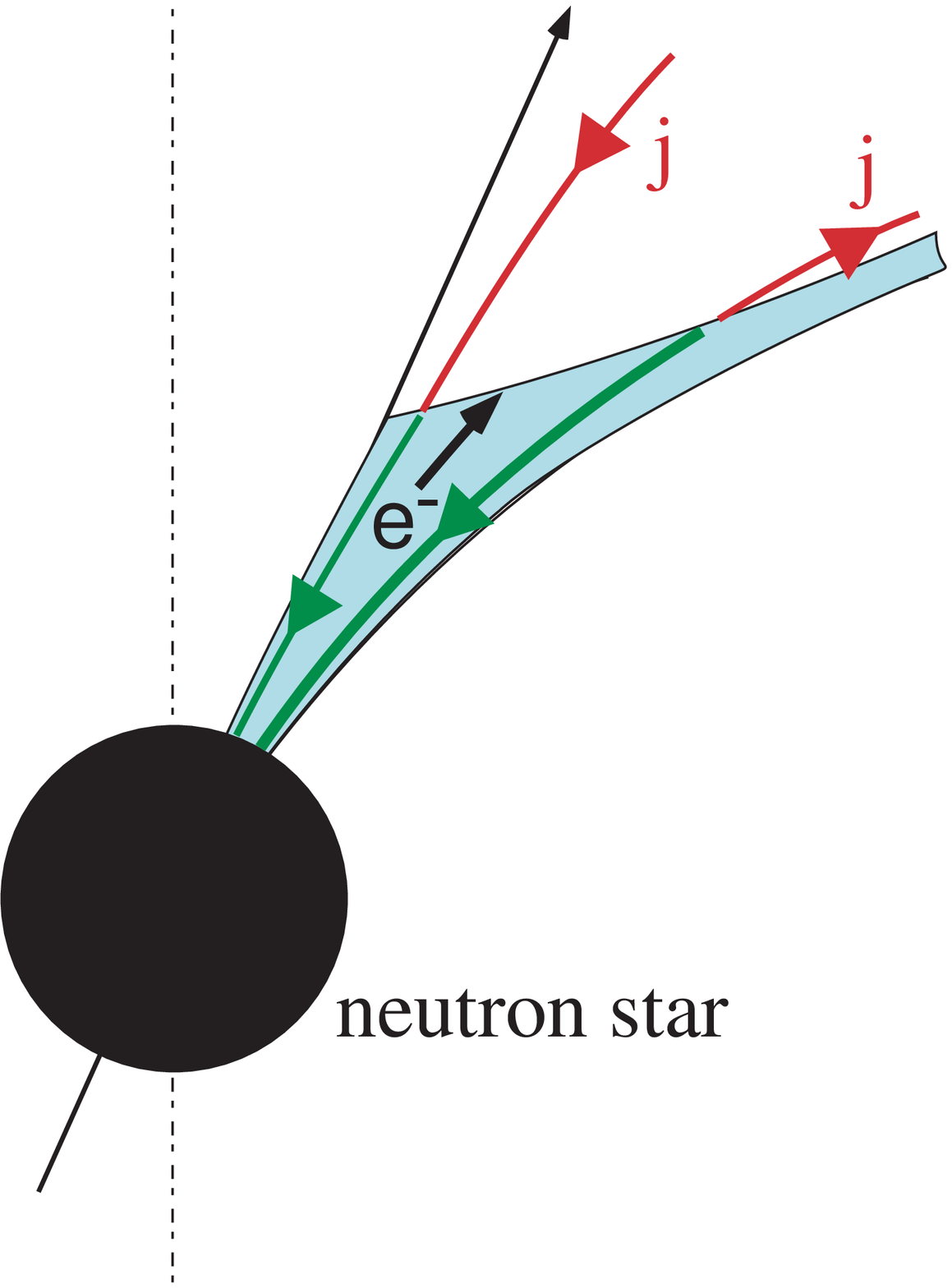}
\caption{
Schematic picture of electric current in the pulsar magnetosphere.
{\it Left}: Global electric current due to the EMF
            exerted on the spinning neutron star surface 
            when $\mbox{\boldmath$\Omega$}\cdot\mbox{\boldmath$\mu$}>0$.
{\it Right}: Current (downward arrows) derived 
            in the inner-slot-gap (shaded region). 
\label{fig:current}
}
\end{figure}

On the spinning neutron star surface, 
an electro-motive force,
$\mbox{EMF}\approx \Omega^2 B_\ast r_\ast^3/c^2 \approx 10^{16.5}$~V,
is exerted from the magnetic pole to the rim of the polar cap.
In this paper, we assume that both the spin and magnetic axes
reside in the same hemisphere; that is, 
$\mbox{\boldmath$\Omega$}\cdot\mbox{\boldmath$\mu$}>0$,
where $\mbox{\boldmath$\Omega$}$ represents the rotation vector,
and $\mbox{\boldmath$\mu$}$ the stellar magnetic moment vector.
This strong EMF causes the magnetospheric currents 
that flow outwards in the lower latitudes
and inwards near the magnetic axis 
(left panel in fig.~\ref{fig:current}).
The return current is formed at large-distances 
where Poynting flux is converted into kinetic energy of particles
or dissipated (Shibata~1997).

Attempts to model the particle accelerator 
have traditionally concentrated on two scenarios:
Polar-cap models with emission altitudes of $\sim 10^4$cm
to several neutron star radii over a pulsar polar cap surface 
(Harding, Tademaru, \& Esposito 1978; Daugherty \& Harding 1982, 1996;
 Dermer \& Sturner 1994; Sturner, Dermer, \& Michel 1995),
and outer-gap models with acceleration occurring in the open
zone located near the light cylinder
(Cheng, Ho, \& Ruderman 1986a,b, hereafter CHR86a,b;
 Chiang \& Romani 1992, 1994; Romani and Yadigaroglu 1995).
Both models predict that electrons and positrons are
accelerated in a charge depletion region, a potential gap,
by the electric field along the magnetic field lines
to radiate high-energy $\gamma$-rays via the curvature 
and inverse-Compton (IC) processes.

In the outer magnetosphere picture of Romani (1996),
he estimated the evolution of
high-energy flux efficiencies and beaming fractions
to discuss the detection statistics,
by considering how pair creation on thermal surface flux
can limit the acceleration zones.
Subsequently, Cheng, Ruderman and Zhang (2000, hereafter CRZ00)
developed a three-dimensional outer magnetospheric gap model,
self-consistently limiting the gap size by pair creation 
from collisions of thermal photons from the polar cap 
that is heated by the
bombardment of gap-accelerated charged particles.
The outer gap models of these two groups 
have been successful in explaining the observed light curves,
particularly in reproducing the wide separation of the two
peaks commonly observed from $\gamma$-ray pulsars
(Kanbach 1999; Thompson 2001),
without invoking a very small inclination angle.
In these outer gap models,
they consider that the gap extends from the null surface,
where the Goldreich-Julian charge density vanishes,
to the light cylinder,
beyond which the velocity of a co-rotating plasma would 
exceed the velocity of light,
adopting the vacuum solution of the Poisson equation for
the electrostatic potential (CHR86a).

However, it was analytically demonstrated by 
Hirotani, Harding, and Shibata (2003, HHS03)
that an active gap, which must be non-vacuum, 
possesses a qualitatively different properties from
the vacuum solution discussed in traditional outer-gap models.
For example, the gap inner boundary shifts towards the star
as the created current increases
and at last touch the star if the created current exceeds
the Goldreich-Julian (GJ) value at the surface.
Therefore, 
to understand the particle accelerator,
which extends from the stellar surface to the outer magnetosphere,
we have to merge the outer-gap and polar-cap models,
which have been separately considered so far.

In traditional polar-cap models,
the energetics and pair cascade spectrum have had success
in reproducing the observations.
However, the predicted beam size of radiation emitted 
near the stellar surface
is too small to produce the wide pulse profiles that are observed
from high-energy pulsars.
Seeking the possibility of a wide hollow cone of high-energy radiation
due to the flaring of field lines,
Arons (1983) first examined the particle acceleration
at the high altitudes along the last open field line.
This type of accelerator, or the slot gap, 
forms because the pair formation front (PFF),
which screens the accelerating electric field, $\Ell$,
in a width comparable to the neutron star radius,
occurs at increasingly higher altitude 
as the magnetic colatitude
approaches the edge of the open field region
(Arons \& Scharlemann 1979).
Muslimov and Harding (2003, hereafter MH03) extended this argument
by including two new features:
acceleration due to space-time dragging, and
the additional decrease of $\Ell$ at the edge of the gap
due to the narrowness of the slot gap.
Moreover, Muslimov and Harding (2004a,b, hereafter MH04a,b) 
matched the high-altitude slot gap solution for $\Ell$
to the solution obtained at lower altitudes (MH03),
and found that the residual $\Ell$ is 
small and constant,
but still large enough at all altitudes to maintain 
the energies of electrons, 
which are extracted from the star, above 5~TeV.

It is noteworthy that the polar-slot gap model proposed by MH04a,b
is an extension of the polar-cap model
into the outer magnetosphere,
assuming that the plasma flowing in the gap consists of only one sign of 
charges. 
This assumption is self-consistently satisfied in their model,
because pair creation in the extended slot gap occurs at a
reduced rate and the pair cascade due to inward-migrating particles
does not take place.
In the polar-slot gap model, 
the completely charge-separated, space-charge-limited flow (SCLF)
leads to a negative $\Ell$ 
for $\mbox{\boldmath$\Omega$}\cdot\mbox{\boldmath$\mu$}>0$.
However, we should notice here 
that the electric current induced by the negative $\Ell$
(right panel of figure~\ref{fig:current})
contradicts with the global current patterns 
(left panel of figure~\ref{fig:current}),
which is derived by the EMF exerted on the spinning neutron-star surface,
if the gap is located near the last-open field line.
(Note that the return current sheet is not assumed 
 on the last-open field line in the slot gap model.)

On these grounds, we are motivated by the need to contrive 
an accelerator model that predicts a consistent current direction
with the global requirement. 
To this aim, it is straightforward to extend recent outer-gap models,
which predict opposite $\Ell$ to polar-cap models,
into the inner magnetosphere.
Extending the one-dimensional analysis along the field lines 
in several outer-gap models
(Hirotani and Shibata 1999a,~b,~c; HHS03), 
Takata, Shibata, and Hirotani (2004, hereafter TSH04)
and Takata et al. (2006, hereafter TSHC06) 
solved the Poisson equation for the electrostatic potential
on the two-dimensional poloidal plane,
and revealed that the gap inner boundary is located inside of the
null surface owing to the pair creation within the gap,
assuming that the particle motion immediately saturates
in the balance between electric and radiation-reaction forces.

In the present paper, we extend TSH04 and TSHC06
by solving the particle energy distribution explicitly,
and by considering a super-GJ current solution with ion emission
from the neutron star surface.
In \S~\ref{sec:basic_eqs}, we formulate the basic equations 
and boundary conditions.
We then apply it to the Crab pulsar in \S~\ref{sec:appl},
and compare the solution with MH04 in \S~\ref{sec:summary}.

\section{Gap Electrodynamics}
\label{sec:basic_eqs}
In this section, we formulate the basic equations to describe the
particle accelerator, extending the method first proposed by 
Beskin et al. (1992) for black-hole magnetospheres.

\subsection{Background Geometry}
\label{sec:spacetime}
Around a rotating neutron star with angular frequency $\Omega$,
mass $M$ and moment of inertia $I$,
the background space-time geometry is given by
(Lense \& Thirring 1918)
\begin{equation}
  ds^2= g_{tt} dt^2 + 2g_{t\varphi}dtd\varphi
       +g_{rr} dr^2 + g_{\theta\theta} d\theta^2
       +g_{\varphi\varphi} d\varphi^2,
  \label{eq:metric_1}
\end{equation}
where
\begin{equation}
  g_{tt}       
    \equiv \left( 1-\frac{r_{\rm g}}{r}\right) c^2, \,
  g_{t\varphi} 
    \equiv ac\frac{r_{\rm g}}{r}\sin^2\theta,
  \label{eq:metric_2}
\end{equation}
\begin{equation}
  g_{rr}   \equiv -\left( 1-\frac{r_{\rm g}}{r}\right)^{-1}, \,
  g_{\theta\theta}   \equiv -r^2, \,
  g_{\varphi\varphi} \equiv -r^2 \sin^2\theta;
  \label{eq:metric_3}
\end{equation}
$r_{\rm g}\equiv 2GM/c^2$ indicates the Schwarzschild radius,
and $a \equiv I\Omega/(Mc)$ parameterizes the stellar angular momentum;
second and higher order terms in the expansion of
$a/r_{\rm g}$ are neglected.
At radial coordinate $r$, the inertial frame is dragged
at angular frequency 
\begin{equation}   
  \omega 
  \equiv -\frac{g_{t\varphi}}{g_{\varphi\varphi}}
  = \frac{I}{M r_\ast^2}
           \frac{r_{\rm g}}{r_\ast}
           \left(\frac{r_\ast}{r}\right)^3 \Omega
  = 0.15 \Omega I_{45} r_6{}^{-3}
  \label{eq:omega}
\end{equation}
where $r_\ast$ represents the stellar radius,
$I_{45} \equiv I/10^{45} \mbox{ erg cm}^2$, and 
$r_6 \equiv r/10\,\mbox{km}$.

\subsection{Poisson Equation for Electrostatic Potential}
\label{sec:poisson}
The first kind equation we have to consider is the
Poisson equation for the electrostatic potential,
which is given by the Gauss's law as
\begin{equation}
  \nabla_\mu F^{t\mu}
  = \frac{1}{\sqrt{-g}}
    \partial_\mu \left[ \frac{\sqrt{-g}}{\rhowSQR}
                       g^{\mu\nu}(-g_{\varphi\varphi}F_{t\nu}
                               +g_{t\varphi}F_{\varphi\nu})
               \right]
  = \frac{4\pi}{c^2} \rho,
  \label{eq:Poisson_1}
\end{equation}
where $\nabla$ denotes the covariant derivative,
the Greek indices run over $t$, $r$, $\theta$, $\varphi$;
$\sqrt{-g}= \sqrt{g_{rr}g_{\theta\theta}\rhowSQR}=cr^2\sin\theta$ and
\begin{equation}
  \rhowSQR \equiv g_{t\varphi}^2-g_{tt}g_{\varphi\varphi}
           = c^2 \left(1-\frac{r_{\rm g}}{r}\right) r^2 \sin^2\theta.
  \label{eq:rhow2}
\end{equation}
If there is an ion emission from the stellar surface
into the magnetosphere,
the total real charge density $\rho$ is given by
\begin{equation}
  \rho= \rho_{\rm e}+\rho_{\rm ion},
\end{equation}
where $\rho_{\rm e}$ denotes the sum of positronic and electronic
charge densities,
while $\rho_{\rm ion}$ does the ionic one.
The six independent components of the field-strength tensor
give the electromagnetic field observed by a distant static
observer (not by the zero-angular-momentum observer) such that
(Camenzind 1986a, b)
\begin{equation}
  E_r=F_{rt}, \, E_\theta=F_{\theta t}, \,
  E_\varphi=F_{\varphi t},
\end{equation}
\begin{equation}
  B^r= \frac{g_{tt}+g_{t\varphi}\Omega}{\sqrt{-g}} F_{\theta\varphi},
  \,
  B^\theta= \frac{g_{tt}+g_{t\varphi}\Omega}{\sqrt{-g}} F_{\varphi r},
  \,
  B_\varphi= -\frac{\rhowSQR}{\sqrt{-g}} F_{r \theta},
  \label{eq:def_EB}
\end{equation}
where $F_{\mu\nu} \equiv A_{\nu,\mu}-A_{\mu,\nu}$ 
and $A_{\mu,\nu}$ denotes the vector potential $A_\mu$
partially differentiated with respect to $x^\nu$.
The strength of the poloidal component of 
the magnetic field is defined as
\begin{equation}
  B_{\rm p} 
  \equiv c^2 \frac{\sqrt{-g_{rr}(B^r)^2-g_{\theta\theta}(B^\theta)^2}}
                  {g_{tt}+g_{t\varphi}\Omega}.
  \label{eq:def_Bp}
\end{equation}

Assuming that the electromagnetic fields are unchanged
in the corotating frame, we can introduce the 
non-corotational potential $\Psi$
such that
\begin{equation}
  F_{\mu t}+\Omega F_{\mu \varphi}
  = -\partial_\mu \Psi(r,\theta,\varphi-\Omega t),
  \label{eq:def_Psi}
\end{equation}
where $\mu= t,r,\theta,\varphi$.
If $F_{A t}+\Omega F_{A \varphi}=0$ holds for $A=r,\theta$,
the angular frequency $\Omega$ of a magnetic field is
conserved along the field line.
On the neutron-star surface,
we impose $F_{\theta t}+\Omega F_{\theta\varphi}=0$ (perfect conductor) 
to find that the surface is equi-potential,
that is, 
$\partial_\theta \Psi= \partial_t \Psi +\Omega \partial_\varphi \Psi=0$
holds. 
However, in a particle acceleration region,
$F_{A t}+\Omega F_{A \varphi}$ deviates from $0$
and the magnetic field does not rigidly rotate
(even though the deviation from the uniform rotation 
 is small when the potential drop in the gap is much less than the EMF
 exerted on the spinning neutron star surface).
The deviation is expressed in terms of $\Psi$, which gives
the strength of the acceleration electric field 
that is measured by a distant static observer as
\begin{equation}
  \Ell \equiv \frac{\mbox{\boldmath$B$}}{B}
              \cdot \mbox{\boldmath$E$}
       = \frac{B^i}{B}(F_{it}+\Omega F_{i\varphi})
       = \frac{\mbox{\boldmath$B$}}{B}
              \cdot (-\nabla\Psi),
  \label{eq:def_Ell}
\end{equation}
where the Latin index $i$ runs over spatial coordinates
$r$, $\theta$, $\varphi$, and an identity 
$B^r F_{r\varphi}+B^\theta F_{\theta\varphi}=0$ is used.

Substituting equation~(\ref{eq:def_Psi}) into (\ref{eq:Poisson_1}),
we obtain the Poisson equation for the
non-corotational potential,
\begin{equation}
  -\frac{c^2}{\sqrt{-g}}
   \partial_\mu 
      \left( \frac{\sqrt{-g}}{\rhowSQR}
             g^{\mu\nu} g_{\varphi\varphi}
             \partial_\nu \Psi
      \right)
  = 4\pi(\rho-\rhoGJ),
  \label{eq:Poisson_2}
\end{equation}
where the general relativistic Goldreich-Julian charge density
is defined as
\begin{equation}
  \rhoGJ \equiv 
      \frac{c^2}{4\pi\sqrt{-g}}
      \partial_\mu \left[ \frac{\sqrt{-g}}{\rhowSQR}
                         g^{\mu\nu} g_{\varphi\varphi}
                         (\Omega-\omega) F_{\varphi\nu}
                 \right].
  \label{eq:def_GJ}
\end{equation}
Using
$g^{rr}=1/g_{rr}$, 
$g^{\theta\theta}=1/g_{\theta\theta}$, 
$g^{r\theta}=g^{\theta r}=0$,
$g^{tt}=-g_{\varphi\varphi}/\rhowSQR$,
$g^{\varphi t}=g_{t\varphi}/\rhowSQR$, and 
$g^{\varphi\varphi}=-g_{tt}/\rhowSQR$,
taking the limit $r \gg r_{\rm g}$, and
noting that 
$ \partial_r     (r\sqrt{-g_{\theta\theta}}B^\theta)
 -\partial_\theta( \sqrt{-g_{rr}}B^r)$
gives the toroidal component of 
$\mbox{\boldmath$\nabla$}\times\mbox{\boldmath$B$}$,
we find that equation~(\ref{eq:def_GJ}) reduces to the
ordinary, special-relativistic expression 
of the Goldreich-Julian charge density
(Goldreich and Julian 1969; Mestel 1971).

Instead of ($r$,$\theta$,$\varphi$),
we adopt in this paper the magnetic coordinates 
($s$,$\theta_\ast$,$\varphi_\ast$)
such that $s$ denotes the distance along a magnetic field line,
$\theta_\ast$ and $\varphi_\ast$ the magnetic colatitude and 
the magnetic azimuth
of the point where the field line intersects the stellar surface.
Defining that $\theta_\ast=0$ corresponds to the magnetic axis
and that $\varphi_\ast=0$ to the plane on which both 
the rotation and the magnetic axes reside,
we can compute spherical coordinate variables as follows:
\begin{equation}
  r(s,\theta_\ast,\varphi_\ast)
  = r_\ast +\int_0^s \frac{B^r(s',\theta,\varphi-\Omega t)}
                          {B(s',\theta,\varphi-\Omega t)} ds',
  \label{eq:def_r}
\end{equation}
\begin{equation}
  \theta(s,\theta_\ast,\varphi_\ast)
  = \theta(0,\theta_\ast,\varphi_\ast) 
   +\int_0^s \frac{B^\theta(s',\theta,\varphi-\Omega t)}
                               {B(s',\theta,\varphi-\Omega t)} ds',
  \label{eq:def_theta}
\end{equation}
\begin{equation}
  \varphi(s,\theta_\ast,\varphi_\ast)-\Omega t
  = \varphi_\ast
   +\int_0^s \frac{B^\varphi(s',\theta,\varphi-\Omega t)}
                  {B(s',\theta,\varphi-\Omega t)} ds',
  \label{eq:def_phi}
\end{equation}
where $\theta(0,\theta_\ast,\varphi_\ast)$ satisfies
$ \sin\theta(0,\theta_\ast,\varphi_\ast)\cdot
  \cos\varphi_\ast\sin\inc
 +\cos\theta(0,\theta_\ast,\varphi_\ast)\cdot
  \cos\inc
 =\cos\theta_\ast $;
$\inc$ represents the angle between the rotation and magnetic axes.
We can numerically compute the transformation matrix
$\partial x^i   / \partial x^{j'}$ and its inverse
$\partial x^{i'}/ \partial x^j$ from 
equations~(\ref{eq:def_r})--(\ref{eq:def_phi}), where
$x^1=r$, $x^2=\theta$, $x^3=\varphi$,
$x^{1'}=s$, $x^{2'}=\theta_\ast$, and $x^{3'}=\varphi_\ast$.
Substituting 
\begin{equation}
  \frac{\partial}{\partial x^i}
  = \frac{\partial s}{\partial x^i}
    \frac{\partial  }{\partial s  }
   +\frac{\partial\theta_\ast}{\partial x^i}
    \frac{\partial  }{\partial\theta_\ast}
   +\frac{\partial\varphi_\ast}{\partial x^i}
    \frac{\partial  }{\partial\varphi_\ast}
\end{equation}
into equation~(\ref{eq:Poisson_2}),
and utilizing $\partial/\partial t= -\Omega \partial/\partial\varphi$,
we obtain the following form of Poisson equation,
which can be applied to arbitrary magnetic field configurations:
\begin{eqnarray}
  &&
  -\frac{c^2 g_{\varphi\varphi}}{\rhowSQR}
  \left( g^{ss}\partial_s^2 
        +g^{\theta_\ast \theta_\ast} \partial_{\theta_\ast}^2
        +g^{\varphi_\ast \varphi_\ast}
            \partial_{\varphi_\ast}^2
        +2g^{s\theta_\ast} \partial_s \partial_{\theta_\ast}
        +2g^{\theta_\ast \varphi_\ast}
            \partial_{\theta_\ast} \partial_{\varphi_\ast}
        +2g^{\varphi_\ast s}
            \partial_{\varphi_\ast} \partial_s
  \right) \Psi
  \nonumber\\
  && -\left( A^s \partial_s
            +A^{\theta_\ast} \partial_{\theta_\ast}
            +A^{\varphi_\ast} \partial_{\varphi_\ast}
      \right) \Psi
  = 4\pi (\rho-\rhoGJ),
  \label{eq:BASIC_1}
\end{eqnarray}
where (see Appendix for explicit expressions)
\begin{eqnarray}
  g^{i'j'} &=& g^{\mu\nu}\frac{\partial x^{i'}}{\partial x^\mu}
                         \frac{\partial x^{j'}}{\partial x^\nu}
  \nonumber\\
           &=& g^{rr}\left(\frac{\partial x^{i'}}{\partial r}
                     \right)_{\theta,\varphi}
                     \left(\frac{\partial x^{j'}}{\partial r}
                     \right)_{\theta,\varphi}
              +g^{\theta\theta}
                     \left(\frac{\partial x^{i'}}{\partial \theta}
                     \right)_{\varphi,r}
                     \left(\frac{\partial x^{j'}}{\partial \theta}
                     \right)_{\varphi,r}
              -\frac{k_0}{\rhowSQR}
                     \left(\frac{\partial x^{i'}}{\partial\varphi}
                     \right)_{r,\theta}
                     \left(\frac{\partial x^{j'}}{\partial\varphi}
                     \right)_{r,\theta}
  \label{eq:def_mag1}
\end{eqnarray}
and
\begin{equation}
  A^{i'} \equiv
  \frac{c^2}{\sqrt{-g}}
   \left\{ \partial_r \left[ \frac{g_{\varphi\varphi}}{\rhowSQR}
                             \sqrt{-g}g^{rr}
                             \left(\frac{\partial x^{i'}}
                                        {\partial r}\right)_{\theta,\varphi}
                      \right]
          +\partial_\theta
                      \left[ \frac{g_{\varphi\varphi}}{\rhowSQR}
                             \sqrt{-g}g^{\theta\theta}
                             \left(\frac{\partial x^{i'}}
                                        {\partial \theta}\right)_{\varphi,r}
                      \right]
    \right\}
   -\frac{c^2 g_{\varphi\varphi}}{\rhowSQR}
    \frac{k_0}{\rhowSQR}
    \left(\frac{\partial^2 x^{i'}}{\partial\varphi^2}\right)_{r,\theta}.
  \label{eq:def_mag2}
\end{equation}
The light surface, 
which is a generalization of the light cylinder,
is obtained by setting
$k_0 \equiv g_{tt}+2g_{t\varphi}\Omega+g_{\varphi\varphi}\Omega^2$
to be zero (e.g., Znajek~1977; Takahashi et al.~1990).
It follows from equation~(\ref{eq:def_Ell}) that
the acceleration electric field is given by
$\Ell = -(\partial\Psi/\partial s)_{\theta_\ast,\varphi_\ast}$.

Let us briefly consider equation~(\ref{eq:BASIC_1})
near the polar cap surface of a nearly aligned rotator.
Since $s \approx r-r_\ast$, $\theta \ll 1$, and 
$B_\varphi B^\varphi \ll B^2$, we obtain
(Scharlemann, Arons \& Fawley~1978, hereafter SAF78;
 Muslimov \& Tsygan~1992, hereafter MT92)
\begin{equation}
  -\frac{1}{r^2}\frac{\partial}{\partial r}
   \left(r^2\frac{\partial\Psi}{\partial r}\right)
  -\frac{1}{r^2(1-r_{\rm g}/r)}
   \left(\frac{\partial\theta_\ast}{\partial\theta}\right)_{\varphi,r}^2
   \left[ \frac{1}{\theta_\ast}
           \frac{\partial}{\partial\theta_\ast}
           \left(\theta_\ast \frac{\partial\Psi}{\partial\theta_\ast}
           \right)
           +\frac{1}{\theta_\ast^2}
            \frac{\partial^2\Psi}{\partial\varphi_\ast^2}
   \right]
    =4\pi(\rho-\rhoGJ).
   \label{eq:Poisson_4}
\end{equation}
Noting that the solid angle element in the metric
of magnetic coordinates is given by
(to the lowest order in $\theta^2$),
\begin{equation}
  g_{\theta_\ast \theta_\ast}d\theta_\ast^2
 +g_{\varphi_\ast \varphi_\ast}d\varphi_\ast^2
 = r_\ast^2 \frac{B(0,\theta_\ast,\varphi_\ast)}
                 {B(s,\theta_\ast,\varphi_\ast)}
   \left( d\theta_\ast^2+\sin^2\theta_\ast d\varphi_\ast^2 
   \right),
\end{equation}
we find that the factor 
\begin{equation}
  g^{\theta_\ast \theta_\ast}
  = \frac{1}{r^2}
    \left(\frac{\partial\theta_\ast}{\partial\theta}
    \right)_{\varphi,r}^2
  = \frac{1}{r_\ast^2}
    \frac{B(s,\theta_\ast,\varphi_\ast)}
         {B(0,\theta_\ast,\varphi_\ast)}
\end{equation}
expresses the effect of magnetic field expansion 
in equation~(\ref{eq:Poisson_4}).
In the same manner, 
in the general equation~(\ref{eq:BASIC_1}),
magnetic field expansion effect is 
essentially contained in 
$g^{\theta_\ast\theta_\ast}$,
$g^{\theta_\ast\varphi_\ast}$,
$g^{\varphi_\ast\varphi_\ast}$,
or equivalently, in the coefficients of the 
second-order trans-field derivatives.
In what follows, we assume that the azimuthal dimension is large
compared with the meridional dimension
and neglect $\varphi_\ast$ dependences.

\subsection{Particle Boltzmann Equations}
\label{sec:Boltz_part}
The second kind equations we have to consider is the 
Boltzmann equations for particles.
At time $t$, position $\mbox{\boldmath$r$}$, 
and momentum $\mbox{\boldmath$p$}$, 
the distribution function $N_+$ of positrons
(or $N_-$ of electrons)
obeys the following Boltzmann equation,
\begin{equation}
  \frac{\partial{N_\pm}}{\partial t}
    + \mbox{\boldmath$v$} \cdot \mbox{\boldmath$\nabla$} N_\pm
    + \left( q\mbox{\boldmath$E$}
             +\frac{\mbox{\boldmath$v$}}{c}
              \times\mbox{\boldmath$B$}
      \right) \cdot 
      \frac{\partial N_\pm}{\partial \mbox{\boldmath$p$}}
  = S_\pm (t,\mbox{\boldmath$r$},\mbox{\boldmath$p$}),
  \label{eq:boltz_1}
\end{equation}
where $\mbox{\boldmath$v$} \equiv $\mbox{\boldmath$p$}$/(m_{\rm e}\Gamma)$;
$m_{\rm e}$ refers to the rest mass of the electron,
$q$ the charge on the particle,
and $\Gamma \equiv 1/\sqrt{1-(\vert \mbox{\boldmath$v$} \vert /c)^2}$
the Lorentz factor.
In a pulsar magnetosphere,
the collision term $S_+$ (or $S_-$) consists of the terms 
representing the appearing and disappearing 
rates of positrons (or electrons) at $\mbox{\boldmath$r$}$
and $\mbox{\boldmath$p$}$ per unit time per unit phase-space volume
due to pair creation, pair annihilation, 
and the energy transfer due to 
IC scatterings and synchro-curvature process.

Imposing a stationary condition 
\begin{equation}
  \frac{\partial}{\partial t} 
  + \Omega \frac{\partial}{\partial \phi} = 0,
  \label{eq:stationary}
\end{equation}
utilizing $\mbox{\boldmath$\nabla$}\cdot\mbox{\boldmath$B$}=0$,
and introducing dimensionless particle densities per unit magnetic flux tube
such that $n_\pm = N_\pm / (\Omega B/2\pi ce)$,
we can reduce the particle Boltzmann equations as
\begin{equation}
  c\cos\chi \frac{\partial n_\pm}{\partial s}
  +\frac{dp}   {dt}\frac{\partial n_\pm}{\partial p}
  +\frac{d\chi}{dt}\frac{\partial n_\pm}{\partial \chi}
  = S_\pm,
 \label{eq:BASIC_2}
\end{equation}
where the upper and lower signs correspond to the
positrons (with charge $q=+e$) and 
electrons ($q=-e$), respectively; 
$p \equiv \vert\mbox{\boldmath$p$}\vert$ and
\begin{equation}
  \frac{dp}{dt} \equiv q\Ell\cos\chi -\frac{P_{\rm SC}}{c}
  \label{eq:char1}
\end{equation}
\begin{equation}
  \frac{d\chi}{dt} \equiv -\frac{q\Ell\sin\chi}{p}
                         +c\frac{\partial(\ln B^{1/2})}{\partial s}
                          \sin\chi,
  \label{eq:char2}  
\end{equation}
\begin{equation}
  \frac{ds}{dt}= c \cos\chi.
  \label{eq:char3}  
\end{equation}
(Introduction of the radiation-reaction force, $P_{\rm SC}/c$,
 will be discussed in the next paragraph.)
For outward- (or inward-) migrating particles,
$\cos\chi>0$ (or $\cos\chi<0$).
Since we consider relativistic particles, 
we obtain $\Gamma=p/(m_{\rm e}c)$.
The second term in the right-hand side of equation~(\ref{eq:char2})
shows that the particle's pitch angle evolves due to the
the variation of $B$ (e.g., \S~12.6 of Jackson 1962).
For example, without $\Ell$, inward-migrating particles 
would be reflected by the magnetic mirrors.
Using $n_\pm$, we can express $\rho_{\rm e}$ as
\begin{equation}
  \rho_{\rm e}
  = \frac{\Omega B}{2\pi c}
    \int\!\!\!\!\int \left[ n_+(s,\theta_\ast,\varphi_\ast,\Gamma,\chi)
                           -n_-(s,\theta_\ast,\varphi_\ast,\Gamma,\chi)
                     \right]
    d\Gamma d\chi.
  \label{eq:def_rhoe}
\end{equation}
The radiation-reaction force due to synchro-curvature radiation
is given by
(Cheng \& Zhang 1996; Zhang \& Cheng 1997), 
\begin{equation}
  \frac{P_{\rm SC}}{c}
  = \frac{e^2 \Gamma^4 Q_2}{12 r_{\rm c}} 
    \left( 1+\frac{7}{r_{\rm c}^2 Q_2^2} \right),
  \label{eq:Psc}
\end{equation}
where
\begin{equation}
  r_{\rm c} \equiv 
  \frac{c^2}{(r_{\rm B}+\rho_{\rm c})(c\cos\chi/\rho_{\rm c})^2
             +r_{\rm B} \omega_{\rm B}^2},
\end{equation}
\begin{equation}
  Q_2^2 \equiv 
  \frac{1}{r_{\rm B}}
  \left( \frac{r_{\rm B}^2+\rho_{\rm c}r_{\rm B}-3\rho_{\rm c}^2}
              {\rho_{\rm c}^3}
         \cos^4\chi
        +\frac{3}{\rho_{\rm c}}
         \cos^2\chi
        +\frac{1}{r_{\rm B}}
         \sin^4\chi
  \right),
\end{equation}
\begin{equation}
  r_{\rm B} \equiv \frac{\Gamma m_{\rm e}c^2 \sin\chi}{eB},
  \qquad
  \omega_{\rm B}\equiv \frac{eB}{\Gamma m_{\rm e}c}
\end{equation}
and 
$\rho_{\rm c}$ is the curvature radius of the magnetic field line.
In the limit of $\chi \rightarrow 0$
(or $\rho_{\rm c} \rightarrow \infty$),
equation~(\ref{eq:Psc}) becomes the expression of pure curvature 
(or pure synchrotron) emission.

Let us briefly discuss the inclusion of the
radiation-reaction force, $P_{\rm SC}/c$, in equation~(\ref{eq:char1}).
Except for the vicinity of the star,
the magnetic field is much less than the critical value 
($B_{\rm cr} \equiv 4.41 \times 10^{13}$~G)
so that quantum effects can be neglected in synchrotron radiation.
Thus, we regard the radiation-reaction force, which is continuous,
as an external force acting on a particle.
Near the star, if $\Gamma (B/B_{\rm cr})\sin\chi>0.1$ holds,
the energy loss rate decreases from the classical formula
(Erber et al. 1966).
If $\Gamma (B/B_{\rm cr})\sin\chi>1$ holds very close to the star,
the particle motion perpendicular to the field is quantized
and the emission is described by the transitions between Landau states;
thus, equation~(\ref{eq:char1}) and (\ref{eq:Psc}) breaks down.
In this case, we artificially put $\chi=10^{-20}$,
which guarantees pure-curvature radiation after the particles 
have fallen onto the ground-state Landau level,
avoiding to discuss the detailed quantum effects in the strong-$B$
region.
This treatment will not affect the main conclusions of this paper,
because the gap electrodynamics is governed by the pair creation
taking place not very close to the star.


Collision terms are expressed as
\begin{eqnarray}
  \lefteqn{ S_\pm(s,\theta_\ast,\varphi_\ast,\Gamma,\chi) 
      = -\int_{-1}^{1} d\mu_{\rm c}
         \int_{E_\gamma<\Gamma} dE_\gamma 
            \etaICg(E_\gamma,\Gamma,\mu_{\rm c}) 
            n_\pm(s,\theta_\ast,\varphi_\ast,\Gamma,\chi)}
  \nonumber \\  
  &+& \int_{-1}^{1} d\mu_{\rm c}
      \int_{\Gamma_i>\Gamma} d\Gamma_i \, 
           \etaICe (\Gamma_i, \Gamma, \mu_{\rm c})
        n_\pm(s,\theta_\ast,\varphi_\ast,\Gamma_i,\chi)
  \nonumber \\  
  &+& \int_{-1}^{1} d\mu_{\rm c}
      \int dE_\gamma 
      \left[ \left( \frac{\partial \eta_{\gamma\gamma}
                                   (E_\gamma,\Gamma,\mu_{\rm c})}
                         {\partial \Gamma}
                   +\frac{\partial \eta_{\gamma B}
                                   (E_\gamma,\Gamma,\mu_{\rm c})}
                         {\partial \Gamma}
	     \right)
             \frac{B_\ast}{B}
             g_\pm(\mbox{\boldmath$r$},E_\gamma,\mbox{\boldmath$k$})
      \right],
  \nonumber \\  
   \label{eq:src_1}
\end{eqnarray}
where $\mu_{\rm c}$ refers to the cosine of the collision angle between
the particles and the soft photons for inverse-Compton scatterings (ICS),
between
the $\gamma$-rays and the soft photons for two-photon pair creation, 
and between
the $\gamma$-rays and the local magnetic field lines for 
one-photon pair creation.
The function $g$ represents the $\gamma$-ray distribution function
divided by $\Omega B_\ast / (2\pi ce)$
at energy $E_\gamma$, momentum $\mbox{\boldmath$k$}$ and 
position $\mbox{\boldmath$r$}$,
where $B_\ast$ denotes the polar-cap magnetic field strength.
Here, $\mbox{\boldmath$k$}$ should be understood to represent the photon
propagation direction, because $E_\gamma$ and $\mbox{\boldmath$k$}$
are related with the dispersion relation (see next section).
Since pair annihilation is negligible, 
we do not include this effect in equation~(\ref{eq:src_1}).

If we multiply $d\Gamma$ on both sides of equation~(\ref{eq:src_1}),
the first (or the second) term in the right-hand side
represents the rate of particles 
disappearing from (or appearing into) the energy interval 
$m_{\rm e}c^2 \Gamma$ and $m_{\rm e}c^2 (\Gamma+d\Gamma)$
due to inverse-Compton (IC) scatterings;
the third term does the rate of two-photon and one-photon pair
creation processes.

The IC redistribution function 
$\etaICg(E_\gamma,\Gamma,\mu_{\rm c})$ represents the probability
that a particle with Lorentz factor $\Gamma$ upscatters photons 
into energies between $E_\gamma$ and $E_\gamma+dE_\gamma$
per unit time when the 
collision angle is $\cos^{-1}\mu_{\rm c}$.
On the other hand, $\etaICe(\Gamma_i,\Gamma,\mu_{\rm c})$ describes
the probability that a particle changes Lorentz factor from
$\Gamma_i$ to $\Gamma$ in a scattering.
Thus, energy conservation gives
\begin{equation}
  \etaICe(\Gamma_i,\Gamma_f,\mu_{\rm c}) 
     = \etaICg[(\Gamma_i-\Gamma_f)m_{\rm e}c^2,\Gamma_i,\mu_{\rm c}].
  \label{eq:rel_etaIC}
\end{equation}
The quantity $\etaICg$ is defined by 
the soft photon flux $dF_{\rm s}/dE_{\rm s}$ and 
the Klein-Nishina cross section $\sigma_{\rm KN}$ as follows (HHS03): 
\begin{eqnarray}
  \lefteqn{\etaICg(E_\gamma,\Gamma,\mu_{\rm c}) 
  = (1-\beta\mu_{\rm c})}
  \nonumber \\
  &\times&
      \int_{E_{\rm min}}^{E_{\rm max}} 
         dE_{\rm s} \frac{dF_{\rm s}}{dE_{\rm s}}
       \int_{b_{i-1}}^{b_i} dE_\gamma \frac{d{E_\gamma}'}{dE_\gamma}
       \int_{-1}^{1} d\Omega_\gamma'
            \frac{d\sigma_{\rm KN}'(E_\gamma,\Gamma,\mu_{\rm c})}
                 {d{E_\gamma}' d\Omega_\gamma' }
  \label{eq:def_etaICg_1}
\end{eqnarray}
where $\beta \equiv \sqrt{1-1/\Gamma^2}$ is virtually unity, 
$\Omega_\gamma$ the solid angle of upscattered photon,
the prime denotes the quantities in the electron (or positron) 
rest frame, and $E_\gamma=(b_{i-1}+b_i)/2$.
In the rest frame of a particle,
a scattering always takes place well above the resonance energy.
Thus, the Klein-Nishina cross section
can be applied to the present problem.
The soft photon flux per unit 
photon energy $E_{\rm s}$ [$s^{-1}\mbox{cm}^{-2}$] 
is written as $dF_{\rm s}/dE_{\rm s}$
and is given by the surface blackbody emission
with redshift corrections at each distance from the star.

The differential pair-creation redistribution function is given by 
 \begin{equation}
   \frac{\partial\eta_{\gamma\gamma}}{\partial\Gamma}
   (E_\gamma,\Gamma,\mu_{\rm c})
   = (1-\mu_{\rm c})
       \int_{E_{\rm th}}^\infty dE_{\rm s}
          \frac{dF_{\rm s}}{dE_{\rm s}}
          \frac{d\sgP(E_\gamma,\Gamma,\mu_{\rm c})}{d\Gamma},
   \label{eq:def_etaP_2}
 \end{equation}
where the pair-creation threshold energy is defined by
\begin{equation}
  E_{\rm th} \equiv \frac{2}{1-\mu_{\rm c}}\frac{1}{E_\gamma},
  \label{eq:def_Eth}
\end{equation}
and the differential cross section is given by
\begin{eqnarray}
  \lefteqn{\frac{d\sgP}{d\Gamma}
    = \frac38 \sgT \frac{1-\bCM^2}{E_\gamma}}
  \nonumber\\
  &\times&
    \left[ \frac{1+\bCM^2(2-\mu_{\rm CM}^2)}{1-\bCM^2\mu_{\rm CM}^2}
          -\frac{2\bCM^4(1-\mu_{\rm CM}^2)^2}{(1-\bCM^2\mu_{\rm CM}^2)^2}
    \right];
  \label{eq:def_sgP}
\end{eqnarray}
$\sgT$ refers to the Thomson cross section and
the center-of-mass quantities are defined as
\begin{equation}
  \mu_{\rm CM} 
  \equiv \pm \frac{2\Gamma m_{\rm e}c^2-E_\gamma}{\bCM E_\gamma}, \quad
  \bCM^2 \equiv 1-\frac{2(m_{\rm e}c^2)^2}{(1-\mu_{\rm c})
                E_{\rm s}E_\gamma}.
\end{equation}

Since a convenient form of 
$\partial \eta_{\gamma B}/\partial \Gamma$ is not given
in the literature, 
we simply assume that all the particles are created
at the energy $\Gamma m_{\rm e}c^2=E_\gamma/2$
for magnetic pair creation.
This treatment does not affect the conclusions 
in the present paper.

Let us briefly mention the electric current per magnetic flux tube.
With projected velocities, $c\cos\chi$, along the field lines,
electric current density in units of $\Omega B / (2\pi)$ is given by
\begin{equation}
  j_{\rm gap}(s,\theta_\ast)
  = j_{\rm e}(s,\theta_\ast)+j_{\rm ion}(\theta_\ast),
  \label{eq:jgap}
\end{equation}
where
\begin{equation}
  j_{\rm e} \equiv \!\!\int\!\!\!\!\int (n_- +n_+) \cos\chi \, dpd\chi;
\end{equation}
$j_{\rm ion}$ denotes the current density carried by the ions
emitted from the stellar surface.
Since $dp/dt$ and $d\chi/dt$ in 
equation~(\ref{eq:BASIC_2}) depend on 
momentum variables $p$ and $\chi$,
$j_{\rm e}$ and hence $j_{\rm gap}$
does not conserve along the field line in an exact sense.

Nevertheless, $j_{\rm gap}$ is kept virtually constant for $s$.
This is because most of the particles have relativistic velocities 
projected along the magnetic field lines at each point.  
For example, consider a situation that a pair is created inwardly at 
potion $s= s_1$.  
In this case, the positron will return after migrating a certain distance
(say, $\delta s$, which is positive). 
In $s_1-\delta s < s < s_1$, the positron does not contribute for the
electric current, because both the inward and the outward current cancel
each other in a stationary situation we are dealing with, 
provided that the projected velocity along the field line is relativistic.
Only when the positronic trajectory on ($s$,$p\cos\chi$) space becomes 
{\it asymmetric} with respect to the $p\cos\chi=0$ axis, owing to the 
synchrotron radiation, which is important if $\vert\cos\chi\vert \ll 1$ 
(see fig. 14 of Hirotani \& Shibata~1999a), 
the returning positron has a non-vanishing contribution for the current 
density at $s \sim s_1-\delta s$.
In $s>s_1$, positronic pitch angle is small enough to give a spatially 
constant contribution to the current density (per magnetic flux tube). 
For electrons, it always has an inward relativistic projected velocity 
and hence gives a spatially constant contribution to the current density.
In practice, the contribution of the returning particles with
an asymmetric trajectory around $\chi\sim 90^\circ$
on the current density,
can be neglected when we discuss the $j_{\rm gap}$.
Thus, we can regard that $j_{\rm gap}$ is virtually conserved even though
$dp/dt$ and $d\chi/dt$ have $p$ and $\chi$ dependences.

\subsection{Gamma-ray Boltzmann Equations}
\label{sec:Boltz_gamm}
The third kind equations we have to consider is the 
Boltzmann equation for $\gamma$-rays.
In general, the distribution function $g$ of the $\gamma$-rays
with momentum $\mbox{\boldmath$k$}$ 
obeys the following Boltzmann equation
\begin{equation}
  \frac{\partial g}{\partial t} 
  + c\frac{\mbox{\boldmath$k$}}{\vert\mbox{\boldmath$k$}\vert}\cdot
    \nabla g(t,\mbox{\boldmath$r$},\mbox{\boldmath$k$}) 
  = S_\gamma(t,\mbox{\boldmath$r$},\mbox{\boldmath$k$}),
  \label{eq:Boltz_gam_0}
\end{equation}
where $\vert\mbox{\boldmath$k$}\vert^2 \equiv -k^i k_i$;
$S_\gamma$ is given by
\begin{eqnarray}   
  S_\gamma
  &=& - \int_{-1}^{1} d\mu_{\rm c} 
        \int_1^\infty d\Gamma
           \frac{\partial\etaP(\mbox{\boldmath$r$},\Gamma,\mu_{\rm c})}
                {\partial\Gamma} 
           \cdot g(\mbox{\boldmath$r$},E_\gamma,
                   k^\theta/k^r,k^\varphi/k^r)
  \nonumber \\
  & & + \int_{-1}^{1} d\mu_{\rm c} 
        \int_1^\infty d\Gamma 
           \etaICg(E_\gamma,\Gamma,\mu_{\rm c}) 
           \frac{B}{B_\ast}
           n_\pm(s,\theta_\ast,\varphi_\ast,\Gamma,\chi)
  \nonumber \\
  & & + \int_0^\pi d\chi
        \int_1^\infty d\Gamma 
           \etaSC( E_\gamma,\Gamma,\chi)
           \frac{B}{B_\ast}
           n_\pm(s,\theta_\ast,\varphi_\ast,\Gamma,\chi),
  \label{eq:def_Sg}
\end{eqnarray}   
where $\etaSC$ is the synchro-curvature radiation rate [s${}^{-1}$]
into the energy interval between $E_\gamma$ and $E_\gamma+dE_\gamma$
by a particle migrating with Lorentz factor $\Gamma$,
$\chi$ the pitch angle of particles.
Explicit expression of $\etaSC$ is given by
Cheng and Zhang (1996).

Imposing the stationary condition~(\ref{eq:stationary}),
or equivalently, assuming that $g$ depends on $\varphi$ and $t$ as
$g=g(r,\theta,\varphi-\Omega t,\mbox{\boldmath$k$})$,
we obtain
\begin{equation}
  \left( c\frac{k^\varphi}{\vert\mbox{\boldmath$k$}\vert}
        -\Omega \right)
   \frac{\partial g}{\partial\bar{\varphi}}
 +c\frac{k^r}{\vert\mbox{\boldmath$k$}\vert}
   \frac{\partial g}{\partial r}
 +c\frac{k^\theta}{\vert\mbox{\boldmath$k$}\vert}
   \frac{\partial g}{\partial \theta}
 = S_\gamma(r,\theta,\bar{\varphi},
            c\vert\mbox{\boldmath$k$}\vert,
            k^r,k^\theta,k_\varphi),
  \label{eq:BASIC_3}
\end{equation}
where $\bar{\varphi}=\varphi-\Omega t$.
To compute $k^i$, 
we have to solve the photon propagation in the curved spacetime.
Since the wavelength is much shorter than the
typical system scales, 
geometrical optics gives
the evolution of momentum and position of a photon 
by the Hamilton-Jacobi equations,
\begin{equation}
  \frac{dk_r     }{d\lambda}=-\frac{\partial k_t}{\partial r},
  \quad
  \frac{dk_\theta}{d\lambda}=-\frac{\partial k_t}{\partial \theta}
  \label{eq:HJ_1}
\end{equation}
\begin{equation}
  \frac{dr      }{d\lambda}=\frac{\partial k_t}{\partial k_r      },
  \quad
  \frac{d\theta }{d\lambda}=\frac{\partial k_t}{\partial k_\theta },
  \label{eq:HJ_2}
\end{equation}
where the parameter $\lambda$ is defined so that $cd\lambda$
represents the distance (i.e., line element) along the ray path.
The photon energy at infinity $k_t$ and 
the azimuthal wave number $-k_\varphi$  
are conserved along the photon path
in a stationary and axisymmetric spacetime
(e.g., in the spacetime described by 
 eqs.~[\ref{eq:metric_1}]--[\ref{eq:metric_3}]).
Hamiltonian $k_t$ can be expressed
in terms of $k_r$, $k_\theta$, $k_\varphi$,
$r$, $\theta$ from the dispersion relation $k^\mu k_\mu=0$,
which is a quadratic equation of $k_\mu$ ($\mu=t,r,\theta,\varphi$).
Thus, we have to solve the set of four ordinary differential 
equations~(\ref{eq:HJ_1}) and (\ref{eq:HJ_2})
for the four quantities, 
$k_r$, $k_\theta$, $r$, and $\theta$ along the ray.
Initial conditions at the emitting point are given by
$k^i/\vert\mbox{\boldmath$k$}\vert=\pm B^i/\vert\mbox{\boldmath$B$}\vert$,
where $i=r$, $\theta$, $\varphi$;
the upper (or lower) sign is chosen for the $\gamma$-rays emitted by an
outward- (or inward-) migrating particle.
When a photon is emitted with energy $E_{\rm local}$ 
by the particle of which angular velocity
is $\dot\varphi$, it is related with $k_t$ and $-k_\varphi$ 
by the redshift relation,
$E_{\rm local}= (dt/d\tau)(k_t+k_\varphi \dot\varphi)$,
where $dt/d\tau$ is solved from
$(dt/d\tau)^2(g_{tt}+2g_{t\varphi}\dot{\varphi}
 +g_{\varphi\varphi}\dot{\varphi}^2)=1$.
To express the energy dependence of $g$,
we regard $g$ as a function of $k_t=E_\gamma$
(i.e., observed photon energy).

%

In this paper, in accordance with the two-dimensional analysis
of equations~(\ref{eq:BASIC_1}) and (\ref{eq:BASIC_2}),
we neglect $\bar{\varphi}$ dependence of $g$,
by ignoring the first term in the 
left-hand side of equation~(\ref{eq:BASIC_3}).
In addition, we neglect the aberration of photons 
and simply assume that the $\gamma$-rays do not have angular momenta
and put $k_\varphi=0$.
The aberration effects are important 
when we discuss how the outward-directed $\gamma$-rays will be observed.
However, they can be correctly taken into account
only when we compute the propagation of emitted photons
in the three-dimensional magnetosphere.
Moreover, they are not essential when we investigate the
electrodynamics,
because the pair creation is governed by the specific intensity of
inward-directed $\gamma$-rays,
which are mainly emitted in a relatively inner region of the magnetosphere.
Thus, it seems reasonable to adopt $k_\varphi=0$ 
when we investigate the two-dimensional gap electrodynamics.

We linearly divide the longitudinal distance into $400$ grids
from $s=0$ (i.e., stellar surface) to $s=1.4\rlc$,
and the meridional coordinate into $16$ field lines
from $\theta_\ast=\theta_\ast^{\rm max}$ (i.e., the last-open field line)
to $\theta_\ast=\theta_\ast^{\rm min}$ (i.e., gap upper boundary),
and consider only $\varphi_\ast=0$ plane
(i.e., the field lines threading the stellar surface on the plane
formed by the rotation and magnetic axes).
To solve the particle Boltzmann equations~(\ref{eq:BASIC_2}), 
we adopt the Cubic Interpolated Propagation (CIP) scheme
with the fractional step technique to shift the profile of
the distribution functions $n_\pm$ in the direction of the
velocity vector in the two-dimensional momentum space
(e.g., Nakamura \& Yabe 1999).
To solve the $\gamma$-ray Boltzmann equation (\ref{eq:BASIC_3}), 
on the other hand,
we do not have to compute the advection of $g$ in the momentum space,
because only the spatial derivative terms remain
after integrating over $\gamma$-ray energy bins,
which are logarithmically divided 
from $\beta_1=0.511$~MeV to $\beta_{29}=28.7$~TeV into $29$ bins.
The $\gamma$-ray propagation directions, 
$k^\theta/k^r$, are divided linearly 
into 180 bins every $\Delta\theta_\gamma=2$~degrees. 
Since the specific intensity in $i$th energy bin
at height 
$ \theta_\ast=\theta_\ast^k
 =\theta_\ast^{\rm max}-(k/16)(\theta_\ast^{\rm max}-\theta_\ast^{\rm min})$,
is given by
\begin{equation}
  g_{i,k,\,l}(s)= \frac{c}{\Delta\theta_\gamma \Delta\phi_\gamma}
              \int_{b_{i-1}}^{b_i} 
              g[s,\theta_\ast^k,E_\gamma,(k^\theta/k^r)_l] \, dE_\gamma,
\end{equation}
the observed $\gamma$-ray energy flux at distance $d$ 
is calculated as
\begin{equation}
  F_{i,\,l}= \frac{\Delta y \sum_{k} \Delta z_k \, g_{i,k,\,l}}
                {d^2},
\end{equation}
where $\Delta y$ denotes the azimuthal dimension of the gap
at longitudinal distance $s$ ($=\rlc$ in this paper),
$\Delta z_k$ the meridional thickness between two field lines
with $\theta_\ast=\theta_{\ast k}$ and $\theta_{\ast k+1}$, and
$i=1,2,3,\ldots,28$, $k=1,2,3,\ldots,15$, $l=1,2,3,\ldots,180$. 
To compute the phase-averaged spectrum, we set the azimuthal width
of the $\gamma$-ray propagation direction, 
$\Delta\phi_\gamma$ to be $\pi$~radian.

Equation~(\ref{eq:def_Sg}) describes the $\gamma$-ray absorption and
creation rate within the gap.
However, to compute observable fluxes, we also have to consider
the synchrotron emission by the secondary, tertiary, and higher-generation
pairs that are created outside of the gap.
If an electron or positron is created with
energy $\Gamma_0 m_{\rm e}c^2$ and pitch angle $\chi_0$,
it radiates the following number of $\gamma$-rays 
(in units of $\Omega B_\ast / 2\pi ce$)
in energies between $b_{i-1}$ and $b_i$: 
\begin{equation}
  \frac{d g_i}{dn}
   = \frac{2\pi ce}{\Omega B_\ast}
     \int_{0}^{\infty} dt
     \int_{b_{i-1}}^{b_i}
     \frac{1}{E_\gamma}
     \frac{dW}{dtdE_\gamma}dE_\gamma,
  \label{eq:2ndary_1}
\end{equation}
where
\begin{equation}
  \frac{dW}{dtdE_\gamma}
  = \frac{\sqrt{3}e^3 B \sin\chi_0}{h m_{\rm e}c^2}
    F\left(\frac{E_\gamma}{E_{\rm c}}\right),
  \label{eq:2ndary_2}
\end{equation}
\begin{equation}
  m_{\rm e}c^2 \frac{d\Gamma}{dt}
  = -\frac{2}{3}
     \frac{e^4 B^2 \sin^2\chi_0}{m_{\rm e}^2c^3} \Gamma^2,
  \label{eq:2ndary_3}
\end{equation}
\begin{equation}
  F(x) \equiv x\int_{x}^\infty K_{5/3}(\xi)d\xi;
  \label{eq:2ndary_4}
\end{equation}
$K_{5/3}$ is the modified Bessel function of $5/3$ order,
and $E_{\rm c}\equiv (3h/4\pi)(eB\Gamma^2\sin\chi_i)/(m_{\rm e}c)$
is the synchrotron critical energy at Lorentz factor $\Gamma$.
Substituting equations~(\ref{eq:2ndary_2}) and (\ref{eq:2ndary_3})
into (\ref{eq:2ndary_1}), we obtain
\begin{equation}
  \frac{d g_i}{dn} 
  = \frac{2\pi ce}{\Omega B_\ast}
    \frac{3\sqrt{3}m_{\rm e}^2 c^3}{2heB\sin\chi_0}
    \int_1^{\Gamma_0} \frac{d\Gamma}{\Gamma^2}
    \int_{b_{i-1}/E_{\rm c}}^{b_i/E_{\rm c}} dy
    \int_{y}^\infty K_{5/3}(\xi)d\xi.
  \label{eq:2ndary_5}
\end{equation}
Note that we assume that particle pitch angle is
fixed at $\chi=\chi_0$, because ultra-relativistic particles emit
radiation mostly in the instantaneous velocity direction,
preventing pitch-angle evolution.
Once particles lose sufficient energies, they preferentially lose
perpendicular momentum; nevertheless, such less-energetic particles
hardly emit synchrotron photons above MeV energies.
On these grounds, to incorporate the synchrotron radiation of 
higher-generation pairs created outside of the gap,
we add $\int_1^\infty (dn/d\Gamma_0)(dg_i/dn) d\Gamma_0$
to compute the emission of $\gamma$-rays in 
the energy interval [$b_{i-1}$,$b_i$]
in the right-hand side of equation~(\ref{eq:Boltz_gam_0}),
where $dn/d\Gamma_0$ denotes the particles 
created between position $s$ and $s+ds$
in Lorentz factor interval $[\Gamma_0,\Gamma_0+d\Gamma_0]$.

\subsection{Boundary Conditions}
\label{sec:BDC}
In order to solve the set of partial differential 
equations~(\ref{eq:BASIC_1}), (\ref{eq:BASIC_2}), and
(\ref{eq:BASIC_3})
for $\Psi$, $n_\pm$, and $g$,
we must impose appropriate boundary conditions.
We assume that the gap {\it lower} boundary,
$\theta_\ast=\theta_\ast^{\rm max}$,
coincides with the last open field line,
which is defined by the condition that 
$ \sin\theta\sqrt{-g_{rr}}B^r
 +\cos\theta\sqrt{-g_{\theta\theta}}B^\theta=0$
is satisfied at the light cylinder on the surface $\varphi_\ast=0$.
Moreover, we assume that the {\it upper} boundary coincides with a specific
magnetic field line
and parameterize this field line with 
$\theta_\ast=\theta_\ast^{\rm min}$.
In general, $\theta_\ast^{\rm min}$ is a function of $\varphi_\ast$;
however, we consider only $\varphi_\ast=0$ in this paper.
Determining the upper boundary from physical consideration
is a subtle issue,
which is beyond the scope of the present paper.
Therefore, we treat $\theta_\ast^{\rm min}$ as a free parameter.
We measure the trans-field thickness of the gap with
\begin{equation}
  h_{\rm m} \equiv \frac{\theta_\ast^{\rm max}-\theta_\ast^{\rm min}}
                        {\theta_\ast^{\rm max}}.
  \label{eq:def_hm}
\end{equation}
If $h_{\rm m}=1.0$, it means that the gap exists along 
all the open field lines. 
On the other hand, if $h_{\rm m}\ll 1$, 
the gap becomes transversely thin and
$\theta_\ast$ derivatives dominate in equation~(\ref{eq:BASIC_1}). 
To describe the trans-field structure,
we introduce the fractional height as
\begin{equation}
  h \equiv \frac{\theta_\ast^{\rm max}-\theta_\ast}
                {\theta_\ast^{\rm max}}.
\end{equation}
Thus, the lower and upper boundaries
are given by $h=0$ and $h=h_{\rm m}$, respectively.

The {\it inner} boundary is assumed to be located at the 
neutron star surface.
For the {\it outer} boundary,
we solve the Poisson equation to a large enough distance, 
$s=1.4\rlc$, which is located outside of the light cylinder.
This mathematical outer boundary is introduced
only for convenience in order that the $\Ell$ distribution
inside of the light cylinder 
may not be influenced by the artificially chosen 
outer boundary position when we solve the Poisson equation.
Since the structure of the outer-most part of the magnetosphere 
is highly unknown, 
we artificially set $\Ell=0$ if the distance from the rotation axis, 
$\varpi$, becomes greater than $0.90\rlc$.
Under this artificially suppressed $\Ell$ distribution in $\varpi>0.90\rlc$,
we solve the Boltzmann equations for outward-migrating 
particles and $\gamma$-rays in $0<s<1.4\rlc$.
For inward-migrating particles and $\gamma$-rays, we solve only in
$\varpi<0.9\rlc$.
The position of the mathematical outer boundary ($1.4\rlc$ in this case), 
little affects the results by virtue of the artificial boundary condition,
$\Ell=0$ for $\varpi>0.9\rlc$.
On the other hand, the artificial outer boundary condition,
$\Ell=0$ for $\varpi>0.9\rlc$, affects the calculation of outward-directed 
$\gamma$-rays to some degree;
nevertheless, it little affects the electrodynamics in the 
inner part of the gap ($s<0.5\rlc$), 
which is governed by the absorption of 
inward-directed $\gamma$-rays.

First, to solve the elliptic-type equation~(\ref{eq:BASIC_1}),
we impose $\Psi=0$ on the lower, upper, and inner boundaries.
At the mathematical outer boundary ($s=1.4\rlc$), 
we impose $\partial\Psi/\partial s=0$.
Generally speaking, 
the solved $\Ell=-(\partial\Psi/\partial s)_{s \rightarrow 0}$
under these boundary conditions
does not vanish at the stellar surface.
Let us consider how to cancel this remaining electric field.

For a super-GJ current density in the sense that
$\rho_{\rm e}-\rhoGJ<0$ holds at the stellar surface,
equation~(\ref{eq:BASIC_1}) gives a positive electric field near the star.
In this case, we assume that ions
are emitted from the stellar surface so that the additional positive
charge in the thin non-relativistic region may bring $\Ell$ to zero
(for the possibility of free ejection of ions due to 
 a low work function, see Jones~1985, Neuhauser et al.~1986, 1987).
The column density in the non-relativistic region becomes (SAF78)
\begin{equation}
  \Sigma_{\rm NR}= \frac{1}{2\pi}
                   \sqrt{\frac{c\Omega B_\ast}{q/m} j_{\rm ion}},
  \label{eq:def_NR}
\end{equation}
where $q/m$ represents the charge-to-mass ratio of the ions
and $j_{\rm ion}$ the ionic current density in units of
$\Omega B_\ast/(2\pi)$.
Equating $4\pi\Sigma_{\rm NR}$ to 
$-(\partial\Psi/\partial s)_{s \rightarrow 0}$
calculated from relativistic positrons, electrons and ions,
we obtain the ion injection rate $j_{\rm ion}$
that cancels $\Ell$ at the stellar surface.

For a sub-GJ current density in the sense that
$\rho_{\rm e}-\rhoGJ>0$ holds at the stellar surface,
$\Psi$ increases outwards near the star
to peak around $s=0.02\rlc \sim 0.10\rlc$,
depending on $\inc$ and $\rho_{\rm e}(s=0)$,
then decrease to become negative in the outer magnetosphere.
That is, $-(\partial\Psi/\partial s)_{s \rightarrow 0}<0$ holds
in the inner region of the gap.
In this case, we assume that electrons
are emitted from the stellar surface 
and fill out the region where $\Psi>0$;
thus, we artificially put $\Psi=0$ if $\Psi>0$ appears.
Even though a non-vanishing, positive $\Ell$ is remained
at the inner boundary, which is located away from the stellar surface,
we neglect such details.
This is because the gap with a sub-GJ current density
is found to be inactive and hence less important,
as will be demonstrated in the next section.

Secondly, to solve the hyperbolic-type equations~(\ref{eq:BASIC_2})
and (\ref{eq:BASIC_3}),
we assume that neither positrons nor $\gamma$-rays are injected
across the inner boundary; thus, we impose 
\begin{equation}
  n_+(s^{\rm in},\theta_\ast,\Gamma,\chi) = 0, \quad
  g(s^{\rm in},\theta_\ast,E_\gamma,\theta_\gamma) = 0 
  \label{eq:BD-2}
\end{equation}
for arbitrary $\theta_\ast$, $\Gamma$, $0<\chi<\pi/2$, $E_\gamma$,
and $\cos(\theta_\gamma-\theta_{\rm B})>0$,
where $\theta_{\rm B}$ designates the outward magnetic field direction.
In the same manner, at the outer boundary, we impose
\begin{equation}
  n_-(s^{\rm out},\theta_\ast,\Gamma,\chi) = 0, \quad
  g(s^{\rm out},\theta_\ast,E_\gamma,\theta_\gamma) = 0 
  \label{eq:BD-4}
\end{equation}
for arbitrary $\theta_\ast$, $\Gamma$, $\pi/2<\chi<\pi$, $E_\gamma$,
and $\cos(\theta_\gamma-\theta_{\rm B})<0$.

\section{Application to the Crab Pulsar}
\label{sec:appl}
Since the formulation described in the foregoing section
is generic, 
we specify some of the quantities in \S~\ref{sec:assumptions}
before turning to a closer examination 
in \S\S\ref{sec:subGJ}--\ref{sec:wind}.

\subsection{Assumptions on Magnetic Field and Soft Photon Field}
\label{sec:assumptions}
First, let us specify the magnetic field.
Near the star, we adopt the static (unperturbed by rotation and currents)
dipole solution obtained in the
Schwarzschild space time 
(e.g., MT92, and references therein).
That is, in equations~(\ref{eq:def_EB}), we evaluate
$F_{\theta\varphi}$ and $F_{\varphi r}$ as
\begin{equation}
  \frac{F_{\theta\varphi}}{\sqrt{-g}}
  = f(r)\frac{2\mu}{r^3}\cos\Theta,
\end{equation}
\begin{equation}
  \frac{F_{\varphi r}}{\sqrt{-g}}
  = -\frac{\mu}{r^2}
     \frac{d}{dr}\left[\frac{f(r)}{r}\right] \sin\Theta,
\end{equation}
where $\Theta$ is the angle measured from the magnetic axis, and
\begin{equation}
 f(r) = -3 \left[ \frac{1}{2} \frac{r}{r_{\rm g}}
                 +\left(\frac{r}{r_{\rm g}}\right)^2
                 +\left(\frac{r}{r_{\rm g}}\right)^3
                  \ln\left(1-\frac{r_{\rm g}}{r}\right)
           \right].
  \label{eq:corr2}
\end{equation}

At the high altitudes (but within the light cylinder), 
the open field lines deviate from the static dipole
to be swept back in the opposite direction of the rotation
and bent toward the rotational equator.
There are two important mechanisms that cause the deviation:
Charge flow along the open field lines, and
retardation of an inclined, rotating dipole.
Both of them appear as the first order correction in 
$\varpi/\rlc$ expansion to the static dipole.
To study the former correction, Muslimov and Harding (2005) 
employed the space-charge-limited longitudinal current solved by MT92, 
and derived an analytic solution of the correction.
However, if we discard the space-charge-limited-flow 
of emitted electrons
and consider copious pair creation in the gap,
we have to derive a more general correction formula that is applicable for
arbitrary longitudinal current distribution.
To follow up this general issue further would involve us in
other factors than the electrodynamics of the accelerator,
and would take us beyond the scope of this paper.
Thus, we consider only the latter correction
and adopt the inclined, vacuum magnetic field solution
obtained by CRZ00 (their equations [B2]--[B4]).

Secondly, we consider how the {\it toroidal} current density,
$J^{\hat\varphi}$, affects $\rhoGJ$ near the light cylinder.
In the outer magnetosphere, general relativistic effects are negligible; 
thus, equation~(\ref{eq:def_GJ}) becomes
\begin{equation}
  \rhoGJ= -\frac{\mbox{\boldmath$\Omega$}\cdot\mbox{\boldmath$B$}}
                {2 \pi c}
          +\frac{\varpi}{\rlc} \frac{J^{\hat\varphi}}{c}.
\end{equation}
Since $J^{\hat\varphi}$ is of the order of $(\Omega B/2\pi)(\varpi/\rlc)$,
the second term appears as a positive correction which is proportional to
$(\varpi/\rlc)^2$
and will become comparable to the first term if $\varpi/\rlc \sim 1$.
Thus, to incorporate this special relativistic correction, we adopt
\begin{equation}
  \rhoGJ= -\frac{\mbox{\boldmath$\Omega$}\cdot\mbox{\boldmath$B$}}
                {2 \pi c}
          \left[ 1+\kappa\left(\frac{\varpi}{\rlc}\right)^2 \right],
  \label{eq:GJ}
\end{equation}
where the constant $\kappa$ is of the order of unity.
For example, CHR86a adopted $\kappa=1$.
Even though a larger value of $\kappa$ is preferable to reproduce
a harder curvature spectrum above 5~GeV and 
a larger secondary synchrotron flux around 100~MeV,
we adopt a conservative value $\kappa=0.5$ in the present paper. 

Thirdly, we have to specify the differential soft photon flux,
$dF_{\rm s}/dE_{\rm s}$,
which appears in equations~(\ref{eq:def_etaICg_1}) and 
(\ref{eq:def_etaP_2}).
As the possible soft photon fields illuminating the gap, 
we can consider the following three components in general:\\
(1) \ 
Photospheric emission from the hot surface of a cooling neutron star.
For simplicity, we approximate this component with 
a black body spectrum with a single temperature, $kT_{\rm s}$.
We assume that this component is uniformly 
emitted from the whole neutron star surface.\\
(2) \
Thermal soft X-ray emission from the neutron star's polar cap
heated by the bombardment of relativistic particles streaming
towards the star from the magnetosphere.
Since we consider a young pulsar in this paper,
this component is negligible compared to the first component.\\
(3) \ 
Non-thermal, power-law emission from charged relativistic particles
created outside of the gap in the magnetosphere.
The emitted radiation can be observed from optical to $\gamma$-ray band.
Since the non-thermal emission will be beamed away from the gap,
we assume that this component does not illuminate the gap.
The major conclusions in this paper will not be affected by this assumption,
except that the pair creation would increase to suppress the 
potential drop and hence the $\gamma$-ray luminosity 
if this component illuminates the gap.
\\

We apply the scheme to the Crab pulsar,
adopting four free parameters,
$\inc$, $\mu$, $kT_{\rm s}$, and $h_{\rm m}$.
Other quantities such as gap geometry 
on the poloidal plane, 
exerted $\Ell$ and potential drop,
particle density and energy distribution, 
as well as the $\gamma$-ray flux and spectrum, 
are uniquely determined if we specify these four parameters.

In the next section, we consider transversely thin and thick cases
in \S~\ref{sec:subGJ} and \S\S~\ref{sec:superGJ}--\ref{sec:wind},
respectively.

\subsection{Sub-GJ current solution: Traditional Outer-gap Model}
\label{sec:subGJ}
To begin with, let us consider the magnetic inclination
$\inc=70^\circ$, 
which is more or less close to the value ($65^\circ$) suggested 
by a three-dimensional analysis in the traditional outer gap model
(CRZ00).
We adopt $kT_{\rm s}=100$~eV as the surface blackbody temperature,
which is consistent with the observational upper limit,
$kT_{\rm s}<180$~eV (Tennant et a.~2001).
In \S\S~\ref{sec:subGJ}--\ref{sec:dep_kT},
we adopt $\mu=4.0 \times 10^{30} \mbox{G cm}^3$
as the magnetic dipole moment,
which gives $B_\ast=1.46 \times 10^{13} \mbox{G}$ (eq.~[\ref{eq:def_Bp}]),
assuming $r_\ast=10^6$~cm and $M=1.4M_\odot$.
If we evaluate $\mu$ from the spin-down luminosity
$\dot{E}=2\Omega^4\mu^2/3c^3$, 
we obtain $\mu=3.8\times 10^{30}\,\mbox{G cm}^3$ for
$\dot{E}=4.46\times 10^{38}\mbox{\,ergs s}^{-1}$
(e.g., Becker \& Tr\"umper~1997).
The dependence of the solution on $kT_{\rm s}$, $\mu$, and $\inc$
will be discussed in \S\S~\ref{sec:dep_kT}--\ref{sec:dep_inc}.

Examine a sub-GJ current solution, 
which is defined by 
$j_{\rm e} 
 < \vert \rhoGJ/(\Omega B/2\pi c) \vert_{s=0}
 \approx 
  (1-\omega/\Omega)B_{z\ast}/B_\ast$,
where $B_{z\ast}$ refers to the surface magnetic field component
projected along the rotation axis;
the right-hand side is evaluated at 
$s=0$ and $\theta_\ast=\theta_\ast^{\rm max}$.
To this aim, we consider a transversely thin gap,
$h_{\rm m}=0.047$.
The solution becomes similar to the vacuum one
obtained in the traditional outer-gap model model (CHR86a),
as the left panel of figure~\ref{fig:E70z047} indicates.
In this figure, we present $\Ell(s,h)$ at discrete height $h$
ranging from $2h_{\rm m}/16$, $5h_{\rm m}/16$, $8h_{\rm m}/16$,
$11h_{\rm m}/16$, $14h_{\rm m}/16$,
with dashed, dotted, solid, dash-dot-dot-dot, and dash-dotted curves,
respectively; they are depicted in the right panel with a
larger $h_{\rm m}$ ($=0.200$) for illustration purpose.
For one thing, for such a transversely thin, nearly vacuum gap, 
the inner boundary is located slightly inside of the null surface.
What is more, $\Ell$ maximizes at the central height, $h=h_{\rm m}/2$,
and remains roughly constant in the entire region of the gap.
The solved $\Ell$ distributes almost symmetrically with respect to the
central height; for example, the dashed and dash-dotted curves
nearly overlap each other.
Similar solutions are obtained for a thiner gap,
$h_{\rm m}<0.047$.

\subsection{Super-GJ current solution: Hybrid gap structure}
\label{sec:superGJ}
\begin{figure}
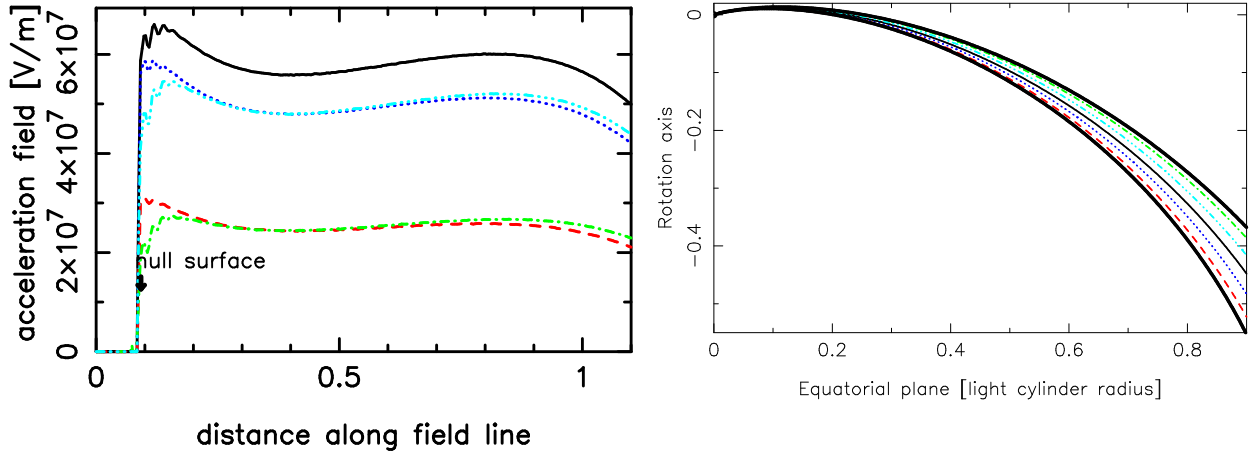

 \includegraphics[angle=-90,scale=1.0]{f03a.ps}
 \hspace*{0.10truecm}
 \includegraphics[angle=-90,scale=0.9]{f03b.ps}
\caption{
{\it Left:}
The field-aligned electric field of a {\it sub}-GJ current solution
at five discrete heights (see right panel)
for $\inc=70^\circ$ and $h_{\rm m}=0.047$.
The abscissa indicates the distance along the field line
from the star in the unit of the light-cylinder radius.
The null surface position at the height $h=h_{\rm m}/2$ 
is indicated by the down arrow.\quad
{\it Right:} Magnetic field lines on the poloidal plane
in which both the rotational and magnetic axes reside.
Instead of $h_{\rm m}=0.047$, $h_{\rm m}=0.200$ is adopted
for clarity.
The thick solid curves denote the lower and upper boundaries,
while the thin dashed, dotted, solid, dash-dot-dot-dotted, and 
dash-dotted ones give the same $h/h_{\rm m}$ values 
as those in the left panel;
they are $h/h_{\rm m}=2/16$, $5/16$, $8/16$, $11/16$, and $14/16$,
respectively.
\label{fig:E70z047}
}
\end{figure}

\begin{figure}
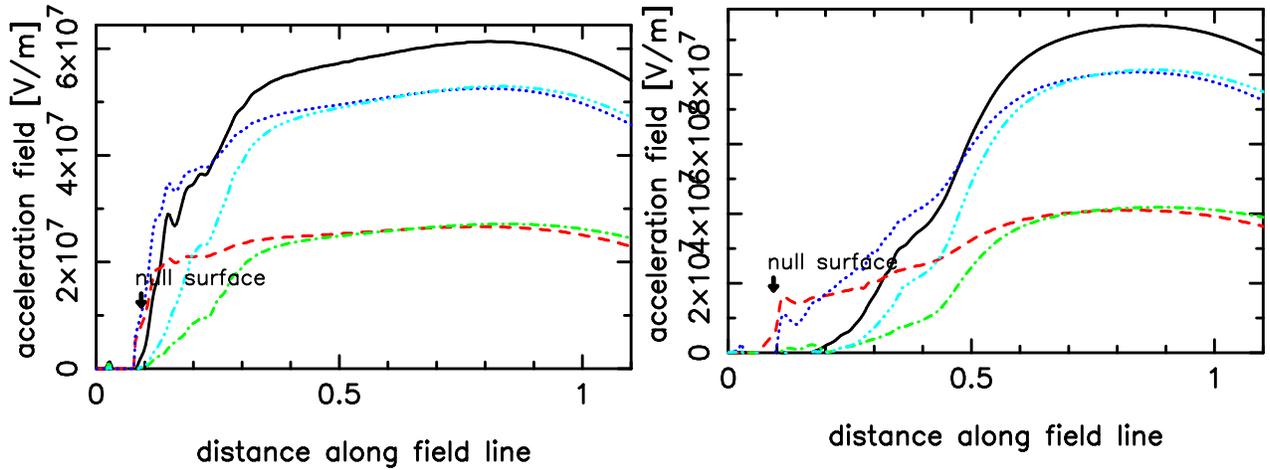

 \includegraphics[angle=-90,scale=1.0]{f04a.ps}
 \includegraphics[angle=-90,scale=1.0]{f04b.ps}
\caption{
Same figure as figure~\ref{fig:E70z047}:
$\Ell(s,h)$ of two {\it super}-GJ current solutions
for $h_{\rm m}=0.048$ (left) and $0.060$ (right)
at five discrete $h$'s.
\label{fig:E70z048-60}
}
\end{figure}

Next, let us consider a thicker gap, $h_{\rm m}=0.048$.
In this case, $j_{\rm e}$ becomes comparable or greater than 
$\vert \rhoGJ/(\Omega B/2\pi c) \vert_{s=0}$ 
in the upper half region $h>h_{\rm m}/2$ of the gap,
deviating the solution from the vacuum one.
In the left panel of figure~\ref{fig:E70z048-60},
we present $\Ell(s,h)$ at five discrete height $h$ 
in the same way as figure~\ref{fig:E70z047}.
It follows that $\Ell$ is screened by the discharge of created pairs
in the inner-most region ($s<0.3\rlc$) in the higher latitudes
($h>h_{\rm m}/2$).
For example, $\Ell$ at $h=7h_{\rm m}/8$ (dash-dotted curve) deviates from
the unscreened solution at the lower latitudes $h=h_{\rm m}/8$ (dashed).
On the other hand, along the lower field lines
($h<h_{\rm m}/2$), $\Ell(s)$ is roughly constant 
as in the traditional outer-gap model.

Let us further consider a thicker gap, $h_{\rm m}=0.060$.
The right panel of figure~\ref{fig:E70z048-60} shows 
that $\Ell(s,h)$ is substantially screened than the
marginally super-GJ case, $h_{\rm m}=0.048$ (left panel).
Because the gap transfield thickness virtually shrinks in $s<0.5\rlc$
due to screening in the higher latitudes,
$\Ell$ in the lower latitudes also decreases
compared to smaller $h_{\rm m}$ cases,
as the dashed lines in the left and right panels indicate.

To understand the screening mechanism, 
it is helpful to examine the Poisson equation~(\ref{eq:Poisson_4}),
which is a good approximation in $s<0.4\rlc$.
In the transversely thin limit, it becomes
\begin{equation}
  -\frac{1}{(1-r_{\rm g}/r)}
   \frac{1}{r_\ast^2}
   \frac{B(r)}{B_\ast}
   \frac{\partial^2\Psi}{\partial\theta_\ast^2}
   \approx
   4\pi(\rho-\rhoGJ)
   \approx
   \frac{2\Omega B(r)}{c}
   \left( \frac{\rho}  {\Omega B/2\pi c}
         -\frac{\rhoGJ}{\Omega B/2\pi c}
   \right).
   \label{eq:Psi_thin}
\end{equation}
Since we are interested in the second-order $\theta_\ast$ derivatives,
this equation is valid not only for a nearly aligned rotator,
but also for an oblique rotator.
We could directly check it from the general equation~(\ref{eq:Poisson_2}).
Factoring out the magnetic field expansion factor,
$B(r)/B_\ast$, from the both sides, we obtain
\begin{equation}
  \Psi
  \approx \left( 1-\frac{r_{\rm g}}{r} \right)
          \frac{\Omega B_\ast}{c}
          \left( \frac{\rho}  {\Omega B/2\pi c} 
                -\frac{\rhoGJ}{\Omega B/2\pi c} 
          \right)
          r_\ast^2 (\theta_\ast-\theta_\ast^{\rm min})
                   (\theta_\ast^{\rm max}-\theta_\ast).
  \label{eq:CHR}
\end{equation}
We thus find that
$\Ell \equiv -\partial\Psi/\partial s$ is approximately proportional to
$-\partial(\rho/B -\rhoGJ/B)/\partial s$. 
It is, therefore,
important to examine the two-dimensional distribution of 
$\rho/B$ and $\rhoGJ/B$ to understand $\Ell(s,h)$ behavior.

\begin{figure}
 \includegraphics[angle=-90,scale=0.65]{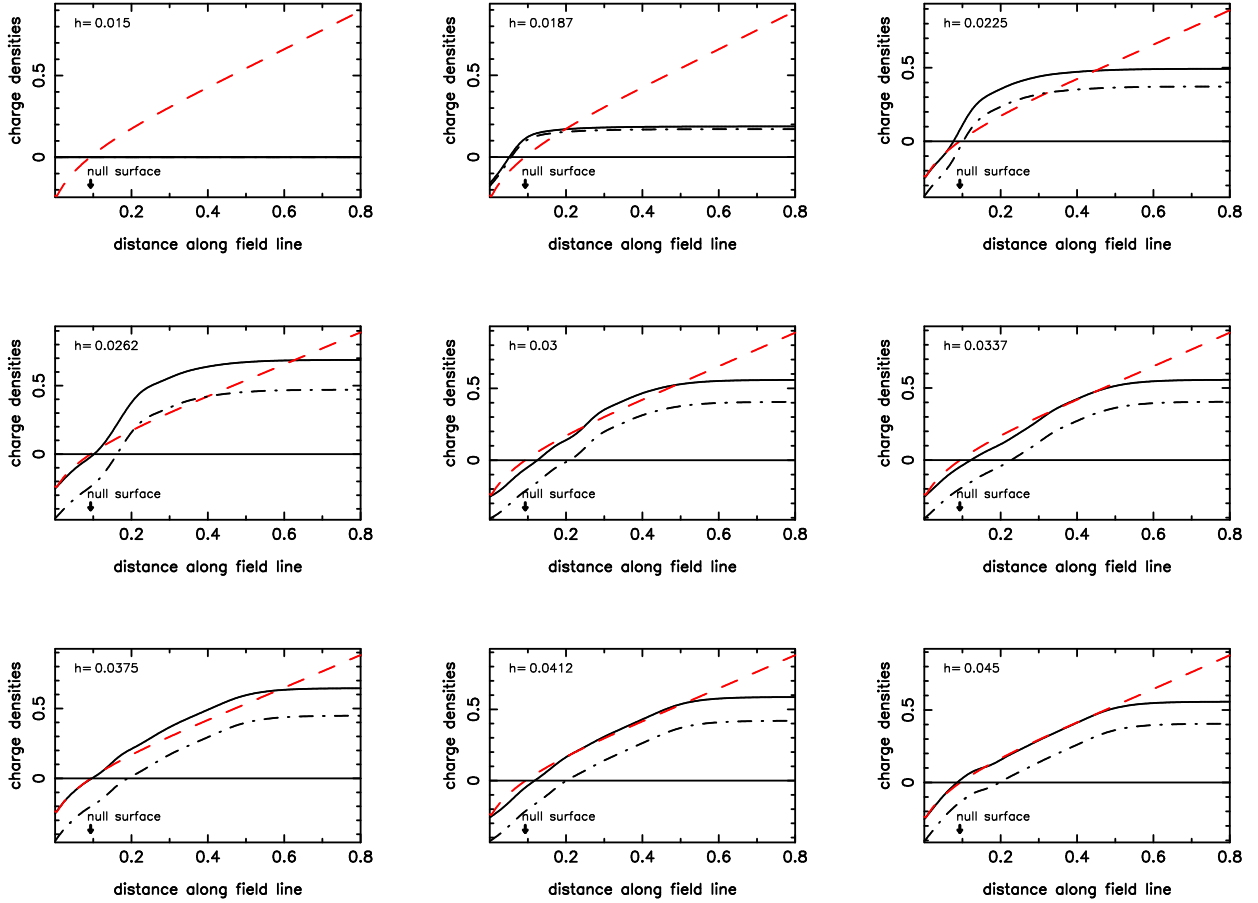}
 \epsscale{.75}
\caption{
Total (solid), created (dash-dotted), and Goldreich-Julian (dashed)
charge densities in $\Omega B(s,h) /(2\pi c)$ unit,
for $\inc=70^\circ$ and $h_{\rm m}=0.060$ 
at nine transfield heights, $h$.
If there is an ion emission from the stellar surface,
the total charge density deviates from the created one.
\label{fig:C70nine}
}
\end{figure}

In figure~\ref{fig:C70nine}, 
we present
$\rho/(\Omega B/2\pi c)$,  
$\rho_{\rm e}/(\Omega B/2\pi c)$, and 
$\rhoGJ/(\Omega B/2\pi c)$,
as the solid, dash-dotted, and dashed curves,
at nine discrete magnetic latitudes, 
ranging from $h=(4/16)h_{\rm m}$, $(5/16)h_{\rm m}$, $\ldots$,
to $(12/16)h_{\rm m}$
for the same parameters as the right panel of figure~\ref{fig:E70z048-60}.
If there is a cold-field ion emission from the star,
the total charge density (solid curve) deviates from
the created charge density (dash-dotted one).
It follows that the current is sub-GJ for $h \le (5/16)h_{\rm m}=0.0187$
and super-GJ for $h \ge (6/16)h_{\rm m}=0.0225$.
Along the field lines with super-GJ current, 
$\rho_{\rm e}-\rhoGJ$ becomes negative close to the star. 
This inevitably leads to a positive $\Ell$,
which extracts ions from the stellar surface.
In this paper, we assume that the extracted ions consist of protons;
nevertheless, the conclusions are little affected by 
the composition of the extracted ions. 

In the outer region, $\rho/B$ levels off 
in $s>0.5\rlc$ for $h>h_{\rm m}/2$.
Since $\rhoGJ/B$ becomes approximately a linear function of $s$,
$\Ell$ remains nearly constant in $s>0.5\rlc$,
in the same manner as in traditional outer-gap model.
In the inner region, on the other hand,
inward-directed $\gamma$-rays propagate into the convex side 
due to the field line curvature,
increasing particle density exponentially with $h$.
As a result, the lower part (i.e., smaller $h$ region) becomes
nearly vacuum.
For example, at $h=h_{\rm m}/4=0.015$ (top left panel), 
positive $\rho_{\rm e}-\rhoGJ$ leads to a negative $\Ell$ 
at the stellar surface, inducing {\it no} ion emission. 
Even though the created current is sub-GJ at $h=0.0187$,
ions are extracted from the surface.
This is because the negative $\rho_{\rm eff} \equiv \rho-\rhoGJ$ 
in the higher latitudes $h \ge 0.0225$
cancels the relatively small positive $\rho_{\rm eff}$ 
along $h=0.0187$ to induce a positive $\Ell$ at the stellar surface.
Such a two-dimensional effect in the Poisson equation is also
important 
in the higher altitudes ($0.1\rlc<s<0.3\rlc$)
along the higher-latitude field lines($h \ge 0.0225$).
Outside of the null surface, $s>0.09\rlc$,
there is a negative $\rho_{\rm eff}$ in the sub-GJ current region
($h \le 0.015$).
This negative $\rho_{\rm eff}$ 
works to prevent $\Ell$ from vanishing in the higher latitudes,
where pair creation is copious.
However, the created pairs discharge until $\Ell$ vanishes,
resulting in a larger gradient of $\rho$
than that of $\rhoGJ$ 
in the intermediate latitudes in 
$0.0225 \le h \le 0.0262$.
In the upper half region ($0.03 \le h < h_{\rm m}=0.06$),
$\partial\rho/\partial s$ does not have to be greater 
than $\partial\rhoGJ/\partial s$, 
in order to screen $\Ell$.

In short, the gap has a hybrid structure:
The lower latitudes (with small $h$) are nearly vacuum 
having sub-GJ current densities
and the inner boundary is located slightly inside of the null surface,
because $\Psi>0$ region will be
filled with the electrons emitted from the stellar surface.
The higher latitudes, on the other hand, are non-vacuum
having super-GJ current densities,
and the inner boundary is located at the stellar surface,
extracting ions at the rate such that their non-relativistic column density
at the stellar surface cancels the strong $\Ell$ 
induced by the negative $\rho-\rhoGJ$ of relativistic electrons, positrons,
and ions.
The created pairs discharge such that
$\Ell$ virtually vanishes in the region where pair creation is copious.
Thus, in the intermediate latitudes between the sub-GJ and super-GJ regions,
$\partial\rho/\partial s > \rhoGJ/\partial s$ holds. 

Even though the inner-most region of the gap is inactive, 
general relativistic effect 
(space-time dragging effect, in this case) is important to determine the
ion emission rate from the stellar surface.
For example, at $h=h_{\rm m}/2$ for $h_{\rm m}=0.600$
(i.e., the central panel in fig.~\ref{fig:C70nine}),
$j_{\rm ion}$ is $69$\% greater than what would be obtained in
the Newtonian limit, 
$\rhoGJ=-\mbox{\boldmath$\Omega$}\cdot\mbox{\boldmath$B$}/2\pi c$.
This is because the reduced $\vert\rhoGJ\vert$ near the star 
(about $15$\% less than the Newtonian value)
enhances the positive $\Ell$,
which has to be canceled by a greater ion emission
(compared to the Newtonian value).
The current, $j_{\rm ion}$, 
is adjusted so that $\vert \rho_{\rm eff} \vert$
may balance with the trans-field derivative of $\Psi$ near the star.
The resultant $\vert \rho_{\rm eff} \vert$
becomes small compared to $\vert\rhoGJ\vert$,
in the same manner as in traditional polar-cap models,
which has a negative $\Ell$ with electron emission from the star.
Although the non-relativistic ions have a large positive charge 
density very close to the star (within $10$~cm from the surface), 
it cannot be resolved in figure~\ref{fig:C70nine}.
Note that the present calculation is performed from the stellar surface
to the outer magnetosphere and does not contain a region with $\Ell<0$.
It follows that an accelerator having $\Ell<0$
(e.g., a polar-cap or a polar-slot-gap accelerator)
cannot exist along the magnetic field lines that have an
super-GJ current density created by the mechanism described in the
present paper.

\begin{figure}
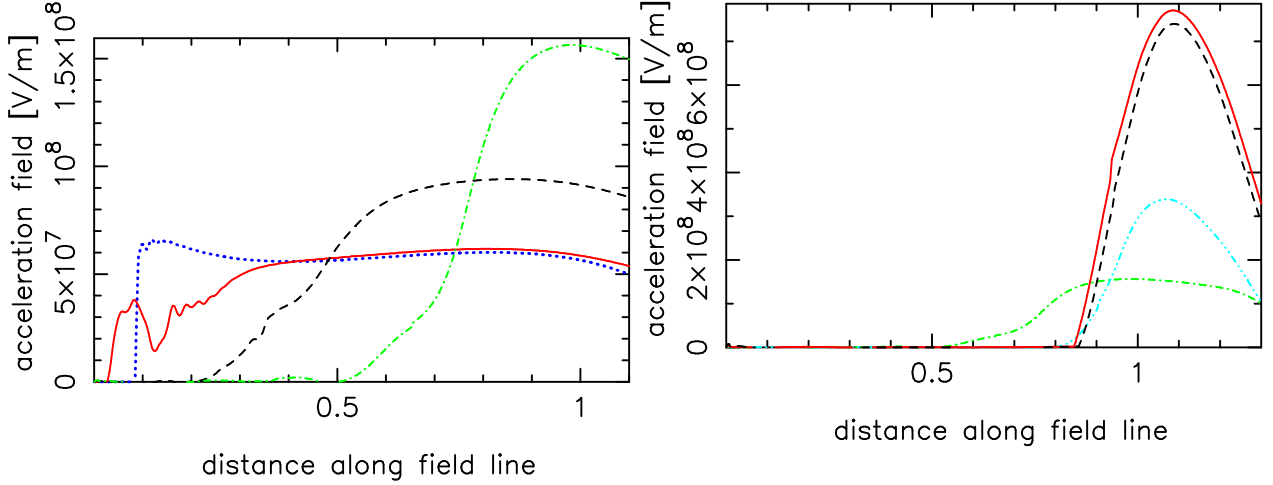

 \includegraphics[angle=-90,scale=1.0]{f06a.ps}
 \includegraphics[angle=-90,scale=1.0]{f06b.ps}
\caption{
Field-aligned electric field at $h=h_{\rm m}/2$ 
as a function of $s/\rlc$
for $\mu=4.0\times 10^{30}\mbox{ G cm}^3$, $kT_{\rm s}=100$~eV,
and $\inc=70^\circ$. 
{\it Left}: 
The dotted, solid, dashed, and dash-dotted curves
corresponds to $h_{\rm}=0.047$, $0.048$, $0.060$, and $0.100$;
{\it right}:
The dash-dotted, dash-dot-dot-dotted, solid, and dashed curves
corresponds to $h_{\rm}=0.100$, $0.160$, $0.200$, and $0.240$.
\label{fig:C70b}
}
\end{figure}

It is worth examining how $\Ell$ changes with varying $h_{\rm m}$.
In figure~\ref{fig:C70b}, 
we present $\Ell(s,h)$ at the central height $h=h_{\rm m}/2$. 
In the left panel, the dotted, solid, dashed, and dash-dotted curves
correspond to $h_{\rm m}=0.047$, $0.048$, $0.060$, and $0.100$, respectively,
while in the right panel,
the dash-dotted, dash-dot-dot-dotted, solid, and dashed ones
to $0.100$, $0.160$, $0.200$, and $0.240$, respectively.
It follows that the inner part of the gap becomes
substantially screened by the discharge of created pairs as 
$h_{\rm m}$ increases.
It also follows that the maximum of $\Ell$ increases with increasing
$h_{\rm m}$ for $h_{\rm m}<0.2$, 
because the two-dimensional screening effect
due to the zero-potential walls becomes less important 
for a larger $h_{\rm m}$.
To solve particle and $\gamma$-ray Boltzmann equations,
we artificially put $\Ell=0$ in $\varpi>0.9\rlc$,
or equivalently in $s>1.1\rlc$ for $\inc=70^\circ$,
as mentioned in \S~\ref{sec:BDC}.

\begin{figure}
 \includegraphics[angle=-90,scale=1.0]{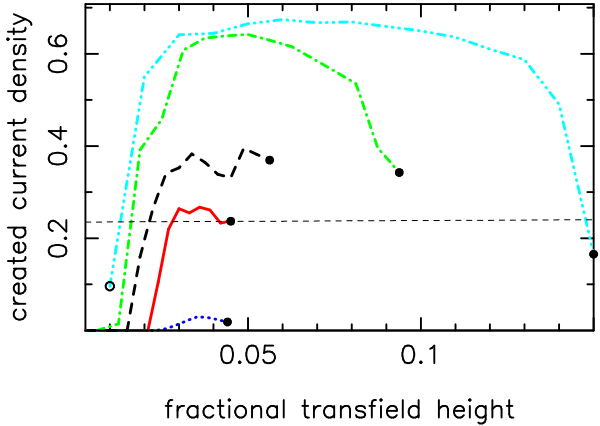}
\caption{
Created current density $j_{\rm e}$ (in unit of $\Omega B/2\pi$)
as a function of the transfield thickness $h$
for $\mu=4.0\times 10^{30}\mbox{ G cm}^3$, $kT_{\rm s}=100$~eV,
and $\inc=70^\circ$.
The dotted, solid, dashed, dash-dotted, and dash-dot-dot-dotted curves
corresponds to 
$h_{\rm}=0.047$, $0.048$, $0.060$, $0.100$, and $0.160$, respectively.
\label{fig:C70d}
}
\end{figure}

The created current density, $j_{\rm e}$, is presented
in figure~\ref{fig:C70d}, as a function of $h$.  
The thin dashed line represents 
$\vert \rhoGJ/(\Omega B/2\pi c) \vert_{s=0}$;
if $j_{\rm e}$ appears below (or above) this line,
the created current is sub- (or super-) GJ along the field line
specified by $h$.
The open and filled circles denote the lowest and highest
latitudes that are used in the computation. 
For $h_{\rm m}=0.047$, the solution (dotted curve) is sub-GJ 
along all the field lines;
thus, screening due to the discharge is negligible as the dotted curve
in the left panel of figure~\ref{fig:C70b} shows.
As $h_{\rm m}$ increases, the solution becomes super-GJ from the
higher latitudes, as indicated by the solid ($h_{\rm m}=0.048$),
dashed ($h_{\rm m}=0.060$), dash-dotted ($h_{\rm m}=0.100$), and
dash-dot-dot-dotted ($h_{\rm m}=0.160$) curves in figure~\ref{fig:C70d}.
As a result, screening becomes significant as $h_{\rm m}$ increase,
as figure~\ref{fig:C70b} shows.
This screening of $\Ell$ has a negative feed back effect 
in the sense that $j_{\rm e}$ is regulated below unity.
Even though it is not resolved in figure~\ref{fig:C70b},
in the lower latitudes, $j_{\rm e}$ grows across the gap height
exponentially, as CHR86a suggested.
For example, $j_{\rm e}=3.0\times 10^{-10}$, 
$2.1\times 10^{-9}$, $2.1\times 10^{-8}$, $1.0\times 10^{-7}$,
$4.3\times 10^{-7}$, $1.0\times 10^{-6}$, $4.6\times 10^{-5}$, and
$1.3\times 10^{-1}$
at $h/h_{\rm m}=1/16$, $2/16$, $3/16$, $\ldots$, $8/16$, respectively,
for $h_{\rm m}= 0.048$ (solid curve).
This is because the pair creation rate at height $h$ 
is proportional to the number of $\gamma$-rays 
that are emitted by charges on all field lines below $h$.

\begin{figure}
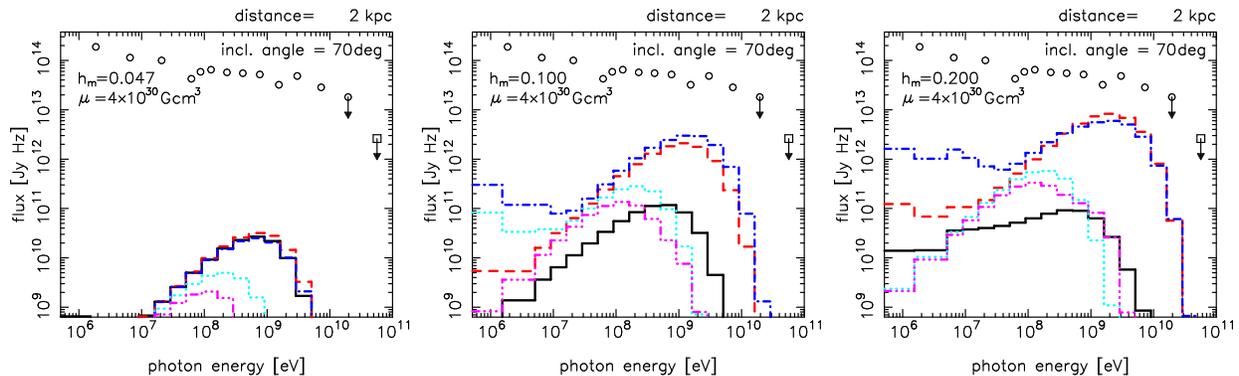

 \includegraphics[angle=-90,scale=0.61]{f08a.ps}
 \includegraphics[angle=-90,scale=0.61]{f08b.ps}
 \includegraphics[angle=-90,scale=0.61]{f08c.ps}
\caption{
Calculated phase-averaged spectra of the 
pulsed, outward-directed $\gamma$-rays
for $\inc=70^\circ$, $kT_{\rm s}=100$~eV, and
$\mu=4.0\times 10^{30} \mbox{ G cm}^3$,
with three different gap thickness, $h_{\rm m}$. 
The flux is averaged over the meridional emission angles between 
$44^\circ$ and $58^\circ$ (solid),
$58^\circ$ and $72^\circ$ (dashed),
$72^\circ$ and $86^\circ$ (dash-dotted), 
$86^\circ$ and $100^\circ$ (dotted), 
$100^\circ$ and $114^\circ$ (dash-dot-dot-dotted),
from the magnetic axis
on the plane in which both the rotational and magnetic axes reside.
\label{fig:C70s1}
}
\end{figure}

Let us now turn to the emitted $\gamma$-rays.
Figure~\ref{fig:C70s1} shows the phase-averaged $\gamma$-ray spectrum
calculated for three different $h_{\rm m}$'s.
Open circles denote the pulsed fluxes detected by
COMPTEL (below 30~MeV; Ulmer et al.~1995, Kuiper et al.~2001) and
EGRET (above 30~MeV; Nolan et al.~1993, Fierro et al.~1998),
while the open square does the upper limit obtained by
CELESTE (de Naurois et al.~2002).
It follows that the sub-GJ (i.e., traditional outer-gap) solution
with $h_{\rm m}=0.047$ predicts too small $\gamma$-ray flux
compared with the observations.
(Note that in traditional outer-gap models, particle number density
is assumed to be the Goldreich-Julian value, while $\Ell$ is given
by the vacuum solution of the Poisson equation, which is inconsistent.)
The maximum flux, which appears around GeV energies, 
does not become greater than $10^{11}$~Jy~Hz for any sub-GJ solutions,
whatever values of $\inc$, $\mu$, and $kT_{\rm s}$ we may choose.
Thus, we can rule out the possibility of a sub-GJ solution
for the Crab pulsar.

As $h_{\rm m}$ increases, the increased $\Ell$ results in a harder 
curvature spectrum,
as the solutions corresponding to $h_{\rm m}=0.100$ and $0.200$ indicate.
As will be discussed in \S~\ref{sec:SR},
the problem of insufficient $\gamma$-ray fluxes may be solved
if we consider a three-dimensional gap structure.
However, the secondary synchrotron flux emitted outside of the gap
is too small to explain the flat spectral shape below 100~MeV.
The $\gamma$-ray spectrum is nearly unchanged for 
$0.2\le h_{\rm m} \le 0.3$.
For $h_{\rm m}>0.3$, the $\gamma$-ray flux tends to decrease,
because the $\Ell(s)$ peaks outside of the artificial outer boundary,
$r\sin\theta=0.9\rlc$, 
which corresponds to $s=1.1\rlc$ for $\inc=70^\circ$.
For $h_{\rm m}>0.4$, the gap virtually vanishes because of the
discharge of copiously created pairs;
in another word, the gap is located outside of $r\sin\theta=0.9\rlc$.
On these grounds, we cannot reproduce the observed flat spectral shape
if we consider $kT_{\rm s}=100$~eV, $\mu=4.0\times 10^{30}\mbox{ G cm}^3$,
and $\inc=70^\circ$,
no matter what value of $h_{\rm m}$ is adopted.
Therefore, in the next three subsections,
we examine how the solution changes if we adopt different
values of $kT_{\rm s}$, $\mu$, and $\inc$.

\subsection{Dependence on Surface Temperature}
\label{sec:dep_kT}
In the same manner as in \S~\ref{sec:superGJ},
we calculate $\Ell(s,h=h_{\rm m}/2)$ for $kT_{\rm s}=150$~eV
to find that their distribution is similar to 
$kT_{\rm s}=100$~eV case (i.e., fig.~\ref{fig:C70b}).
For example, sub/super-GJ current solutions are discriminated 
by the condition whether $h_{\rm m}$ is greater than $0.048$ or not,
and the maximum of $\Ell$ is $7.2 \times 10^8 \mbox{ V m}^{-1}$
for $0.20<h_{\rm m}<0.24$.
Other quantities such as the particle and $\gamma$-ray distribution
functions are also similar.
Therefore, we can conclude that the solution is little subject to change
for the variation of $kT_{\rm s}$,
even though the photon-photon pair production rate increases
with increasing $kT_{\rm s}$.
This is due to the negative feedback effect,
which will be discussed in \S~\ref{sec:stability}.

\subsection{Dependence on Magnetic Moment}
\label{sec:dep_mu}
\begin{figure}
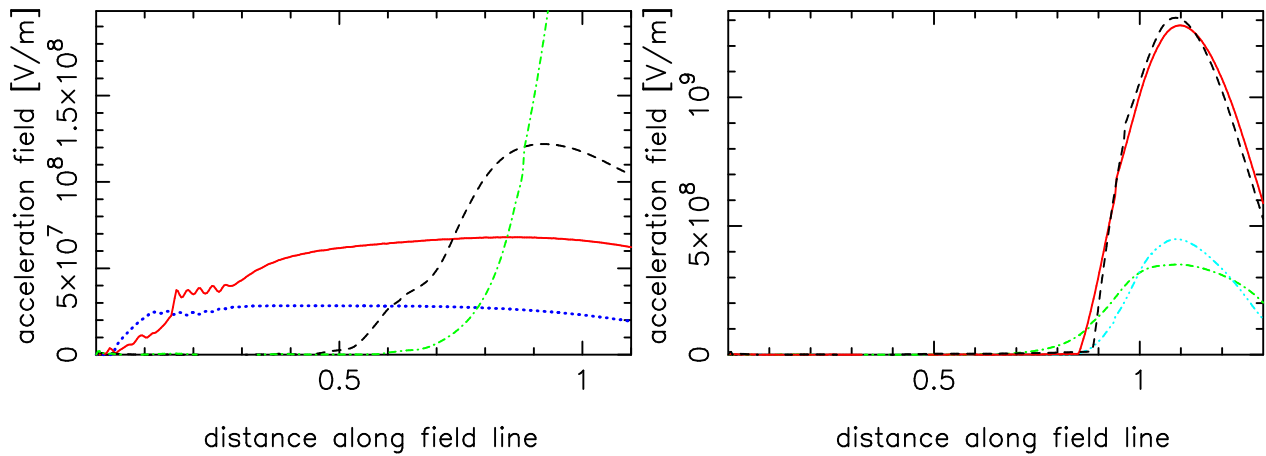

 \includegraphics[angle=-90,scale=1.0]{f09a.ps}
 \includegraphics[angle=-90,scale=1.0]{f09b.ps}
\caption{
Same figure as figure~\ref{fig:C70b}
but with $\mu=6.0\times 10^{30}\mbox{ G cm}^3$;
in the left panel, the dotted and solid curves correspond to 
$h_{\rm}=0.039$ and $0.041$, respectively. 
For other two curves in the left panel and four curves in the right panel,
$h$ takes the same values as in figure~\ref{fig:C70b}.
\label{fig:C70h}
}
\end{figure}

Let us examine how the solution depends on the magnetic moment, $\mu$.
In figure~\ref{fig:C70h},
we present $\Ell(s,h=h_{\rm m}/2)$ for seven discrete $h_{\rm m}$'s
with $\mu=6.0 \times 10^{30} \mbox{ G cm}^3$;
in the left panel, the dotted and solid curves
correspond to $h_{\rm}=0.039$ and $0.041$, respectively,
instead of $h_{\rm}=0.047$ and $0.048$ in figure~\ref{fig:C70b}.
It follows that the exerted $\Ell$ is greater than 
the case of $\mu=4.0\times 10^{30} \mbox{ G cm}^3$ (fig.~\ref{fig:C70b}),
because $\rhoGJ$ increases $1.5$ times.
The negative feedback effect cannot cancel the increase of $\mu$,
unlike the increase of $kT_{\rm s}$,
because the right-hand side of equation~(\ref{eq:BASIC_1})
is more directly affected by the variation of $\mu$ through $\rhoGJ$
than by the variation of $kT_{\rm s}$ through pair creation.

\begin{figure}
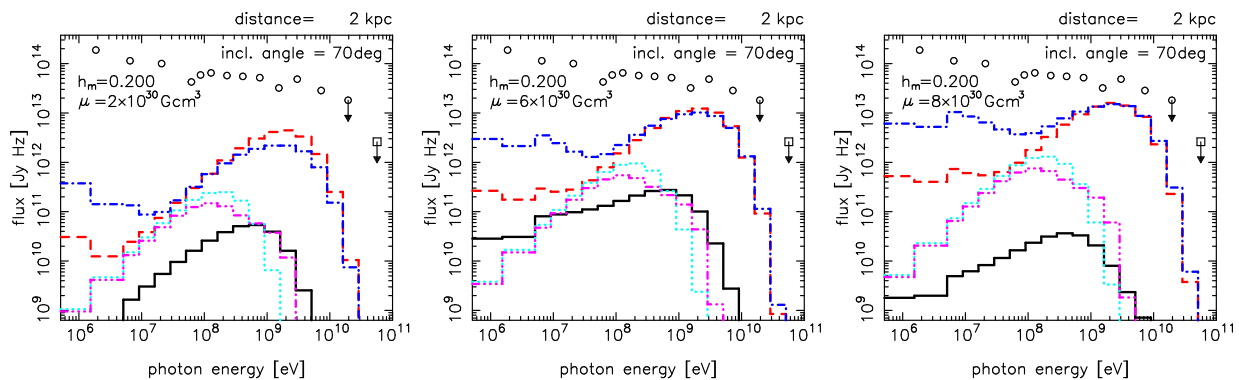

 \includegraphics[angle=-90,scale=0.61]{f10a.ps}
 \includegraphics[angle=-90,scale=0.61]{f10b.ps}
 \includegraphics[angle=-90,scale=0.61]{f10c.ps}
\caption{
Same figure as figure~\ref{fig:C70s1} 
for $\inc=70^\circ$, $kT_{\rm s}=100$~eV, and $h_{\rm m}=0.200$,
with different dipole moment, $\mu$. 
\label{fig:C70s2}
}
\end{figure}

We also present the $\gamma$-ray spectrum for three different
$\mu$'s in figure~\ref{fig:C70s2}.
It follows that both the peak energy and the flux of
curvature $\gamma$-rays (around GeV energies) 
increase with increasing $\mu$.
This is because $\rhoGJ$, and hence $\Ell$ increases with $\mu$.
It also follows that the secondary synchrotron flux
(below 100~MeV) increases with $\mu$,
because the magnetic field strength increases in the magnetosphere.
We find that a larger magnetic dipole moment, 
$\mu \ge 6 \times 10^{30} \mbox{\,G\,cm}^3$ is preferable to explain
the observed pulsed flux from the Crab pulsar.

If we adopt $\mu=8\times 10^{30} \mbox{\,G\,cm}^3$,
which is about twice larger than the dipole deduced value,
the spectral shape becomes more consistent compared with smaller
$\mu$ cases.
We should notice here that the moment of inertia, $I$, have to be large
in this case.
For example, if we assume a pure magnetic dipole radiation,
the spin-down luminosity becomes
$L_{\rm SD}=(2\sin^2\inc/3)(\mu^2 \Omega^4/c^3)
 = 1.7 \times 10^{39} \mbox{\,ergs\ s}^{-1}$
for $\inc=70^\circ$.
Equating $-\dot{E}=-I\Omega\dot\Omega$ with this $L_{\rm SD}$,
we obtain $I=3.9\times 10^{45} \mbox{\,g\ cm}^2$,
which is consistent with the limit 
($I>3.04\times 10^{45}\mbox{\,g\ cm}^2$) 
derived from the consideration of energetics of the Crab nebula
(Bejger \& Haensel~2002).
However, solving the time-dependent equations of force-free electrodynamics,
Spitkovsky (2006) derived
$L_{\rm SD} \approx (1+\sin^2\inc)(\mu^2 \Omega^4/c^3)$,
which gives $\approx 5.6 \times 10^{39} \mbox{\,ergs\ s}^{-1}$
for $\inc=70^\circ$.
This large spin-down luminosity results in 
$I \approx 1.2 \times 10^{46}\mbox{\,g\ cm}^2$, which is too large
even compared with those obtained for stiff equation of state
(Serot~1979a, b; Pandharipande \& Smith~1975).
Thus, it implies either that $\mu$ should be less than 
$8\times 10^{30} \mbox{\,G\,cm}^3$ or that 
the deduced magnetospheric current is too large
when the relationship 
$L_{\rm SD} \approx (1+\sin^2\inc)(\mu^2 \Omega^4/c^3)$
is derived
(for a discussion of the magnetospheric current determination,
 see section 4 of Hirotani~2006).

\subsection{Dependence on Magnetic Inclination}
\label{sec:dep_inc}
\begin{figure}
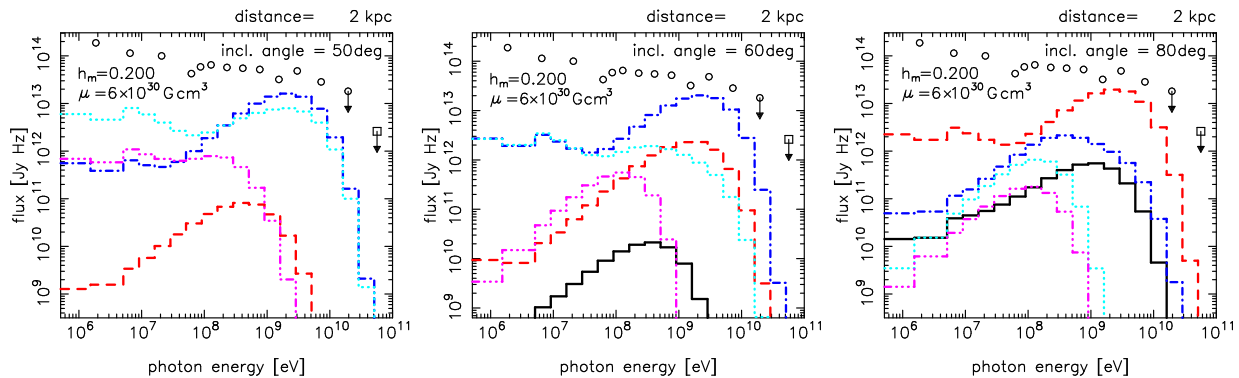

 \includegraphics[angle=-90,scale=0.61]{f11a.ps}
 \includegraphics[angle=-90,scale=0.61]{f11b.ps}
 \includegraphics[angle=-90,scale=0.61]{f11c.ps}
\caption{
Same figure as figure~\ref{fig:C70s1} 
for $kT_{\rm s}=100$~eV, $\mu=6.0\times 10^{30} \mbox{ G cm}^3$, 
and $h_{\rm m}=0.200$
with different magnetic inclination angle, $\inc$.
\label{fig:incl}
}
\end{figure}

Let us examine how the solution depends on the magnetic inclination angle,
$\inc$.
Figure~\ref{fig:incl} shows the $\gamma$-ray spectra 
for $\mu=6\times 10^{30} \mbox{ G cm}^3$, $kT_{\rm s}=100$~eV,
$h_{\rm m}=0.200$, with three different inclination, 
$\inc=50^\circ$, $60^\circ$, and $80^\circ$,
in the same manner as in figure~\ref{fig:C70s1}.
The flux is averaged over the meridional emission angles between 
$44^\circ$ and $58^\circ$ (solid),
$58^\circ$ and $72^\circ$ (dashed),
$72^\circ$ and $86^\circ$ (dash-dotted), 
$86^\circ$ and $100^\circ$ (dotted), 
$100^\circ$ and $114^\circ$ (dash-dot-dot-dotted),
from the magnetic axis
on the plane in which both the rotational and magnetic axes reside.
We find that the $\gamma$-ray flux reaches a peak of 
$\sim 2 \times 10^{13}$~Jy~Hz around 2~GeV 
and that this peak does not strongly depend on $\inc$.
This is because the pair creation efficiency,
which governs the gap electrodynamics,
crucially depends on the distance from the star,
which has a small dependence on $\inc$.
However, we also find that the flux tends to be emitted into larger
meridional angles from the magnetic axis 
(i.e., from outer regions) for smaller $\inc$.
The reasons are fourfold:\\
(1) $\rhoGJ$ decreases with decreasing $\inc$ at a fixed $s$; 
as a result, the null surface appears at larger $s$ for smaller $\inc$.\\
(2) $\rhoGJ(s)=-\rhoGJ(0)$ is realized at larger $s$ for smaller $\inc$.\\
(3) In the region where $\rhoGJ<-\rhoGJ(0)$ holds, 
$\Ell$ is substantially screened by the discharge of created pairs.
(Compare fig.~\ref{fig:C70nine} and 
 the right panel of fig.~\ref{fig:E70z048-60}.)\\
(4) The unscreened $\Ell$ tends to appear at larger $s$ for smaller $\inc$,
resulting in a $\gamma$-ray emission which concentrate in
larger meridional angles.\\
We can alternatively interpret the explanation above as follows:\\
(a) $\vert\rhoGJ(0)\vert$ increases with decreasing $\inc$.\\
(b) The created current density, $j_{\rm e}$, which is greater than 
$c\vert\rhoGJ(0)\vert$ for a super-GJ solution, increases with
decreasing $\inc$.
For example, we obtain $j_{\rm e}=0.62$, $0.67$, $0.74$, and $0.77$
for $\inc=80^\circ$, $70^\circ$, $60^\circ$, and $50^\circ$, respectively,
at $h=h_{\rm m}/2=0.100$
with $\mu=6\times 10^{30} \mbox{ G cm}^3$, and $kT_{\rm s}=100$~eV.\\
(c) A larger $j_{\rm e}$ results in a larger injection
of the discharged positrons and the emitted ions into 
the strong-$\Ell$ region from the stellar side.\\
(d) A larger injection of positive charges from the stellar side
shifts the gap outwards by the mechanism discussed in
\S~2 of Hirotani and Shibata (2001a).

For a smaller inclination, $\inc \le 40^\circ$,
the gap is located in $\varpi>0.9\rlc$;
that is, no super-GJ solution exists.
Therefore, the observed $\gamma$-ray flux cannot be explained 
by the present theory, if the magnetic inclination is constrained to be
less than $40^\circ$ by some other methods.


\subsection{Particle Distribution Functions}
\label{sec:Fn}
\begin{figure}
 \includegraphics[angle=-90,scale=1.0]{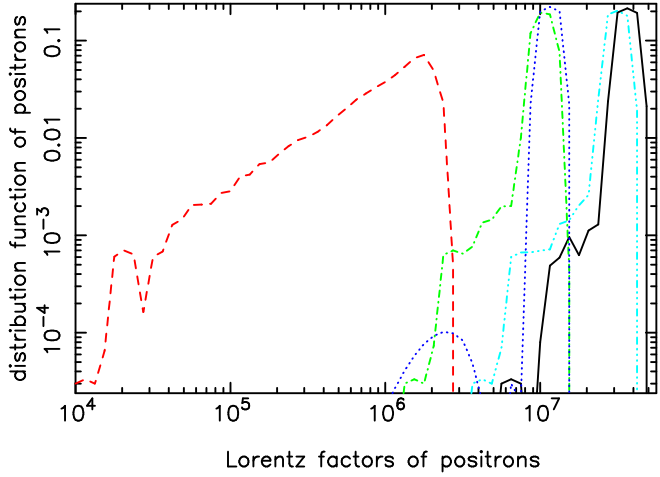}
\caption{
Energy spectrum of positrons at
$s=0.85\rlc$ (dashed), $0.90\rlc$ (dash-dotted), 
$0.95\rlc$ (dash-dot-dot-dotted),
$1.00\rlc$ (solid), and $1.30\rlc$ (dotted),
for $\inc=70^\circ$, 
$\mu=6.0\times 10^{30} \mbox{ G cm}^3$, $kT_{\rm s}=100$~eV,
and $h_{\rm m}=0.20$.
\label{fig:Fn70p1}
}
\end{figure}

Let us examine how the particle distribution function evolves
at different positions.
Figure~\ref{fig:Fn70p1} represents the evolution of positronic
distribution function from 
$s=0.85\rlc$ (dashed curve),
$0.90\rlc$ (dash-dotted),
$0.95\rlc$ (dash-dot-dot-dotted),
$1.00\rlc$ (solid), to
$1.30\rlc$ (dotted).
It shows that the positrons are injected into the strong-$\Ell$ region
($s>0.85\rlc$) with energies $10^4 < \Gamma < 3 \times 10^6$
because of the acceleration by the small-amplitude, residual $\Ell$
in $s<0.85\rlc$ (solid curve in the right panel of fig.~\ref{fig:C70h}). 
The positrons are accelerated outwards to attain
$\Gamma \sim 4 \times 10^7$ at $s \sim 1.0 \rlc$.
They are subsequently decelerated by curvature cooling 
in $\varpi>0.9\rlc$ (or equivalently $s>1.1\rlc$ for $\inc=70^\circ$),
where we artificially put $\Ell=0$,
to escape outwards with $\Gamma \sim 10^7$ at $s \sim 1.3\rlc$.
There is a small population of the positrons that have upscattered 
surface X-rays to possess smaller energies than the
curvature-limited positrons.
For example, within the gap ($s<1.1\rlc$),
the dash-dotted, dash-dot-dot-dotted, and solid curves
have the broad, lower-energy component,
which connects with the curvature-limited peak component.
However, in $s>1.1\rlc$, we artificially put $\Ell=0$;
as a result, the positrons that have lost energies by ICS 
cannot be re-accelerated,
forming a separate component in $\Gamma < 4 \times 10^6$ 
from the curvature-limited peak component, 
as the dotted curve shows. 
The upscattered photons obtain several TeV energies;
however, they are totally absorbed by the strong magnetospheric
infrared radiation field.
Therefore, we depict only the photon energies below $100$~GeV 
in figures~\ref{fig:C70s1}, \ref{fig:C70s2}, and \ref{fig:incl}.

Since most of the pairs are created inwards, positrons return
to migrate outwards by the small-amplitude $\Ell$ in 
$s<0.85\rlc$ 
(for $\mu=6.0\times 10^{30} \mbox{ G cm}^3$ and $\inc=70^\circ$),
losing significant transverse momenta via synchrotron process
to fall onto the ground-state Landau level in strong $B$ region.
Thus, their emission in strong-$\Ell$ region 
is given by a pure-curvature formula.

\begin{figure}
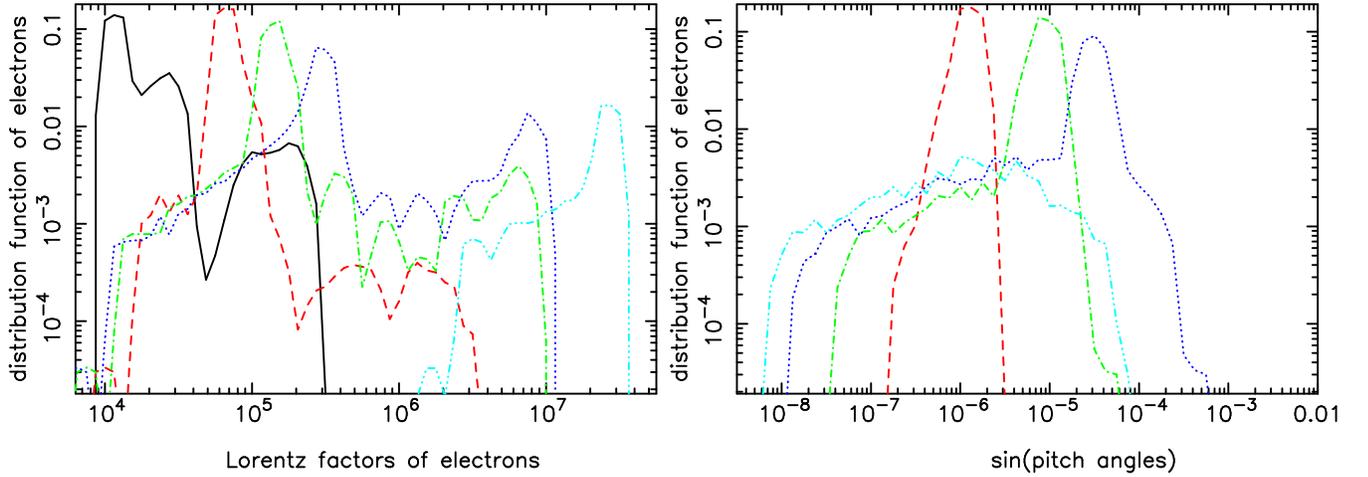

 \includegraphics[angle=-90,scale=1.0]{f13a.ps}
 \includegraphics[angle=-90,scale=1.0]{f13b.ps}
\caption{
Distribution function of electrons at
$s=0.9\rlc$ (dash-dot-dot-dotted), 
$0.4\rlc$ (dotted), 
$0.2\rlc$ (dash-dotted),
$0.08\rlc$ (dashed), and 
$0$ (solid),
for $\inc=70^\circ$, 
$\mu=6.0\times 10^{30} \mbox{ G cm}^3$, $kT_{\rm s}=100$~eV,
and $h_{\rm m}=0.20$.
Left panel shows the Lorentz factor dependence, while the
right one shows the pitch angle dependence.
\label{fig:Fn70e1}
}
\end{figure}

Next, we consider the distribution function of electrons.
In figure~\ref{fig:Fn70e1} we present their evolution 
along the field line $h=h_{\rm m}/2=0.100$ from
$s=0.90\rlc$ (dash-dot-dot-dotted curve),
$0.40\rlc$ (dotted),
$0.20\rlc$ (dash-dotted),
$0.08\rlc$ (dashed), to
$0$ (solid).
Left panel shows the energy spectrum. 
Electrons created in $s<0.85\rlc$ cannot be accelerated by $\Ell$
efficiently; thus, their energy spectrum becomes broad as the
dotted, dashed-dotted, and dashed curves indicate
(i.e., particle Lorentz factors do not concentrate 
 at the curvature-limited terminal value).
From $s=0.08\rlc$ to $s=0$, electrons are decelerated 
via synchro-curvature radiation and 
non-resonant inverse-Compton scatterings.
Finally, they hit the stellar surface with $\Gamma < 3 \times 10^5$.
Assuming that the azimuthal gap width is $\pi$ radian,
we obtain $L_{\rm PC}=2.6\times 10^{31} \mbox{\,ergs s}^{-1}$
as the heated polar-cap luminosity. 
Thus, the X-ray emission due to the bombardment is negligible,
compared with the total soft-X-ray luminosity (0.1--2.4~keV) of
$7.6\times 10^{34} \mbox{\,ergs s}^{-1} 
 (\Delta\Omega_{\rm X}/\mbox{ster})$ 
(e.g., Becker \& Tr\"umper~1997),
where $\Delta\Omega_{\rm X}$ refers to the emission solid angle.

Let us see the pitch angle evolution of the electrons,
which is presented by the right panel of figure~\ref{fig:Fn70e1}.
In the outer part of the gap, 
the electrons are created by the collisions between
the outward-directed $\gamma$-rays and the surface X-rays.
Thus, created electrons have outward momenta initially,
then return by the positive $\Ell$, 
losing their perpendicular momentum substantially via
synchrotron radiation.
Thus, at $s=0.90\rlc$, the dash-dot-dot-dotted curve show
that their pitch angles, $\chi$, are less than $8 \times 10^{-5}$.
However, most of the pairs are created by
the inward-directed $\gamma$-rays in $s<0.6\rlc$;
thus, electrons have initial inward momenta to migrate
inwards by the small-amplitude, residual $\Ell$.
Since such inward-created electrons do not change their 
migration direction,
their pitch angles are greater than those of the outward-created ones,
as the dotted curve demonstrates.
As the electrons migrate inwards,
they lose perpendicular momenta via synchro-curvature radiation
in the strong $B$ field,
reducing their pitch angles,
as the dotted, dash-dotted, and dashed curves indicate.

\begin{figure}
 \includegraphics[angle=-90,scale=1.0]{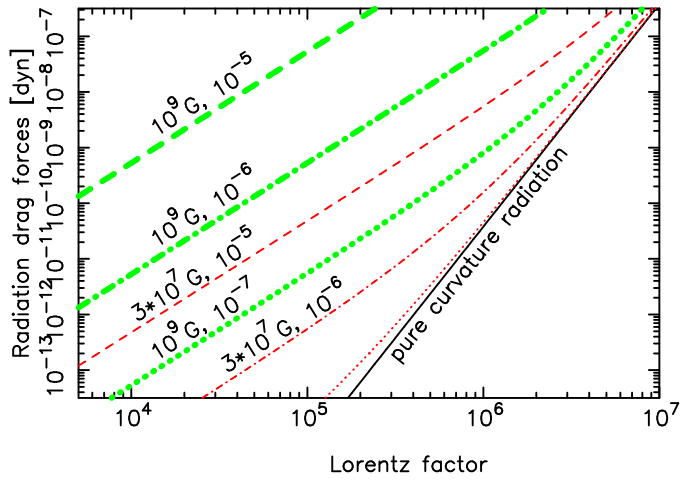}
\caption{
Pure-curvature (solid) vs. synchro-curvature radiation-reaction forces.
For the latter, we adopt the pitch angles 
$\chi=10^{-5}$, $10^{-6}$, and $10^{-7}$~rad 
for the dashed, dash-dotted, and dotted curves, respectively,
and $B=3\times 10^7$ and $10^9$~G for the 
thin and thick ones, respectively. 
Curvature radius is assumed to be $0.4\rlc=6.37\times 10^7$~cm.
\label{fig:DragF}
}
\end{figure}

We should point out that a pure curvature formula cannot be applied
to the electrons.
For example, at $s=0.4\rlc$, the dotted curve 
in figure~\ref{fig:Fn70e1} demonstrates that
electrons have $10^4<\Gamma<10^7$ and $10^{-8}<\sin\chi<10^{-3.5}$.
Noting that we have $B=3.0\times 10^7$~G at $s=0.4\rlc$,
we find that newly created electrons, which have
lower-energies ($\Gamma<10^6$) and larger pitch angles ($\sin\chi>10^{-5}$),
emit synchro-curvature radiation,
rather than pure-curvature one, as figure~\ref{fig:DragF} shows.
At $s=0.2\rlc$, we have $B=3.6\times 10^8$~G; 
thus, the dash-dotted curve shows that 
the pure-curvature formula is totally inapplicable 
as we consider a smaller distance from the star.
The only exception is the inner-most region 
($s<5r_\ast=0.031\rlc$ and $B>5\times 10^{10}$~G),
where electrons suffer substantial de-excitation
via synchrotron radiation, 
falling at last onto the ground-state Landau level.
In this region, electrons emit via pure curvature radiation.
We thus artificially assume $\sin\chi=10^{-20}$, 
which guarantees pure curvature emission,
and do not depict the solid curve 
in the right panel of figure~\ref{fig:Fn70e1}.
Since pair creation and the resultant screening of $\Ell$
is governed by the inward-directed $\gamma$-ray flux and spectrum,
it is essential to adopt the correct radiation formula
by computing the pitch-angle evolution of inward-migrating particles.

\subsection{Formation of Magnetically Dominated Wind}
\label{sec:wind}
Let us finally consider the magnetic dominance within the light cylinder.
First, introduce the magnetization parameter,
\begin{eqnarray}
  \sigma 
  &\equiv &
    \frac{B^2}{4\pi}
    \left\{ \Gamma m_{\rm e}c^2
            \int\!\!\!\!\!\int\!\!\!\!\!\int 
            (N_+ +N_-) d^3\mbox{\boldmath$p$}
           +\Gamma_{\rm ion} m_{\rm ion}c^2
            \int\!\!\!\!\!\int\!\!\!\!\!\int 
            N_{\rm ion} d^3\mbox{\boldmath$p$}
    \right\}^{-1} \nonumber\\
  &=& \frac{e(B/2)\rlc}{ j_{\rm e}  \Gamma           m_{\rm e}c^2
                        +j_{\rm ion}\Gamma_{\rm ion} m_{\rm ion}c^2},
  \label{eq:def_sigma}
\end{eqnarray}
where $N_{\rm ion}$, $\Gamma_{\rm ion}$, refers to the
distribution function and the averaged Lorentz factor of the ions.
Evaluating equation~(\ref{eq:def_sigma}) at the light cylinder,
and noting 
$\Gamma_{\rm ion} m_{\rm ion}c^2=\vert e\Delta\Psi_{\rm gap}\vert$,
where $\vert e\Delta\Psi_{\rm gap}\vert \approx (B/4)h_{\rm m}^2 \rlc$
refers to the potential drop in the gap
(see eq.~[\ref{eq:CHR}]),
we obtain
\begin{equation}
  \sigma_{\rm L} 
    \approx \frac{1}{h_{\rm m}^2} \cdot
            \frac{2}{ j_{\rm e}\Gamma m_{\rm e}c^2
                      /\vert e\Delta\Psi_{\rm gap}\vert
                     +j_{\rm ion}}.
  \label{eq:sigmaL}
\end{equation}
Recalling $j_{\rm ion}<j_{\rm e}<1$, 
$\Gamma m_{\rm e}c^2 < \vert e\Delta\Psi_{\rm gap}\vert$, and
$h_{\rm m}<1$,
we can conclude that the magnetic energy flux is always greater than
the particle kinetic energy flux at the light cylinder,
regardless of the species of the accelerated particles, 
of the sign of $\Ell$,
or of the gap position (i.e., whether inner or outer magnetosphere).
Only when the gap is formed along most of the open field lines 
(i.e., $h_{\rm m}\sim 1$),
$\sigma_{\rm L}$ can be of the order of unity.
Since the second factor in equation~(\ref{eq:sigmaL}) does not change
significantly for different values of parameters except for $h_{\rm m}$,
$\sigma_{\rm L}$ solely depends on $h_{\rm m}$.
Substituting $j_{\rm ion} \sim 0.3$, $j_{\rm e} \sim 0.6$,
$\Gamma m_{\rm e}c^2/\vert e\Delta\Psi_{\rm gap}\vert \sim 0.25$
in $(3/16)h_{\rm m} \le h \le (13/16)h_{\rm m}$
for $\inc=70^\circ$, $\mu=6\times 10^{30}\mbox{ G cm}^3$, and
$h_{\rm m}=0.20$, we obtain $\sigma_{\rm L} \sim 110$. 

In short, along the open field lines threading the gap,
Poynting flux dominates particle kinetic energy flux by a factor of 100
at the light cylinder.
Positrons escape from the gap with $\Gamma \sim 10^7$,
while ions with $\Gamma_{\rm ion} \sim 10^4$.

\section{Summary and Discussion}
\label{sec:summary}
To conclude, we investigated 
the self-consistent electrodynamic structure of a particle accelerator
in the Crab pulsar magnetosphere
on the two-dimensional surface that contains the magnetic field lines 
threading the stellar surface on the plane in which both
the rotation and magnetic axes reside.
We regard the trans-field thickness, $h_{\rm m}$, of the gap as a
free parameter, instead of trying to constrain it.
For a small $h_{\rm m}$, 
the created current density, $j_{\rm e}$, becomes sub-Goldreich-Julian,
giving the traditional outer-gap solution
but with negligible $\gamma$-ray flux.
However, as $h_{\rm m}$ increases, $j_{\rm e}$ increases to become
super-GJ, giving a new gap solution with substantially screened
acceleration electric field, $\Ell$, in the inner part.
In this case, the gap extends towards the neutron star 
with a small-amplitude positive $\Ell$,
which extracts ions from the stellar surface.
It is essential to examine the pitch-angle evolution of the
created particles, because the inward-migrating particles emit
$\gamma$-rays, 
which governs the gap electrodynamics through pair creation,
via synchro-curvature process rather than pure-curvature one.
The resultant spectral shape of the outward-directed $\gamma$-rays 
is consistent with the existing observations;
however, their predicted fluxes appear insufficient.
The pulsar wind at the light cylinder is magnetically dominated:
Along the field lines threading the gap, magnetization parameter,
$\sigma_{\rm L}$, is about $10^2$. 


\subsection{How to obtain sufficient gamma-ray flux}
\label{sec:SR}
The obtained $\gamma$-ray fluxes in the present work, 
are all below the observed values.
Nevertheless, this problem may be solved if we
extend the current analysis into a three-dimensional configuration space.
As mentioned in \S~\ref{sec:Boltz_gamm},
we assume $k_\varphi=0$ and neglect the aberration of photon
emission directions.
However, in a realistic three-dimensional pulsar magnetosphere,
$\gamma$-rays will have angular momenta and may 
be emitted in a limited solid angle as suggested by 
the caustic model (e.g., Dyks \& Rudak~2003; Dyks, Harding, \& Rudak 2004),
which incorporate the effect of aberration of photons
and that of time-of-flight delays.
In particular, in the trailing peak of a highly inclined rotator,
photons emitted at different altitudes $s$ will be beamed in a 
narrow solid angle to be piled up at the same phase of a pulse
(Morini~1983; Romani \& Yadigaroglu~1995),
resulting in a $\gamma$-ray flux which is an order of
magnitude greater than the present values.
Thus, the insufficient $\gamma$-ray fluxes
does not suggest the inapplicability of the present method.
It is noteworthy that the meridional propagation angles of 
the emitted photons (e.g., different curves in figs.~\ref{fig:C70s1},
 \ref{fig:C70s2}, and \ref{fig:incl})
can be readily translated into the emissivity distribution 
in the gap as a function of $s$.
Therefore, in this work, 
we do not sum up the $\gamma$-ray fluxes emitted into different
meridional angles taking account of the aberration of light.

\subsection{Stability of the Gap}
\label{sec:stability}
Let us discuss the electrodynamic stability of the gap,
by considering whether an initial perturbation of 
some quantity tends to be canceled or not.
In the present paper, we consider that the soft photon field
is given and unchanged when gap quantities vary.
Thus, let us first consider the case when the soft photon field is fixed.
Imagine that the created pairs are decreased as an initial perturbation.
It leads to an increase of the potential drop due to less efficient
screening by the discharged pairs,
and hence to an increase of particle energies.
Then the particles emit synchro-curvature radiation efficiently,
resulting in an increase of the created pairs,
which tends to cancel the initial decrease of created pairs.

Let us next consider the case when the soft photon field also changes.
Imagine again that the created pairs are decreased 
as an initial perturbation.
It leads to an increase of particle energies in the same manner as
discussed just above.
The increased particle energies increase 
not only the number and density of synchro-curvature $\gamma$-rays,
but also the surface blackbody emission from heated polar caps
and the secondary magnetospheric X-rays.
Even though neither the heated polar-cap emission nor the 
magnetospheric emission are taken into account 
as the soft photon field illuminating the gap in this paper,
they all work, in general, to increase the pair creation within the gap,
which cancels the initial decrease of created pairs
more strongly than the case of the fixed soft photon field.

Because of such negative feedback effects,
solution exists in a wide parameter space.
For example, the created current density is almost unchanged for
a wide range of $h_{\rm m}$
(e.g., compare the dash-dotted and dash-dot-dot-dotted curves in 
 fig.~\ref{fig:C70d}).
On these grounds,
although the perturbation equations are not solved
under appropriate boundary conditions for the perturbed quantities,
we conjecture that the particle accelerator is electrodynamically stable,
irrespective whether the X-ray field illuminating the gap
is thermal or non-thermal origin.

\subsection{Local vs. global currents}
\label{sec:current}
Let us briefly look at the relationship between the
locally determined current density $j_{\rm gap}$ (eq.~[\ref{eq:jgap}])
and the globally required one, $j_{\rm global}$.
It is possible that $j_{\rm global}$ is constrained
independently from the gap electrodynamics
by the dissipation at large distances (Shibata~1997),
which provides the electric load in the current circuit,
or by the condition that the magnetic flux function should be
continuous across the light cylinder,
as discussed in recent force-free argument of the trans-field equation
(Contopoulos, Kazanas, Fendt 1999; Goodwin et al. 2004;
 Gruzinov 2005; Spitkovsky 2006).
In either case, $j_{\rm global}$ will be more or less close to
unity (i.e., typical GJ value).
On the other hand, as demonstrated in \S~\ref{sec:superGJ},
for super-GJ cases,
$j_{\rm gap}$ is automatically regulated around $0.9$ 
for a wide parameter range.
Thus, the discrepancy between $j_{\rm gap}$ and $j_{\rm global}$ 
is small provided that $h_{\rm m}$ is large enough to maintain
the created current density at a super-GJ value.
The small imbalance $j_{\rm global}-j_{\rm gap}$ may have to be
compensated by a current injection across the outer boundary
(if the gap terminates inside of the light cylinder,
 charged particles could be injected from the outer boundary),
or by an additional ionic emission from the stellar surface
(if the imbalance leads to an additional residual $\Ell$ at the surface).
In any case, the injected current is small compared with $j_{\rm gap}$;
thus, it will not change the electrodynamics significantly,
even though the gap active region may be shifted to some degree
along the magnetic field lines,
as demonstrated by Hirotani and Shibata (2001a, b; 2002), 
TSH04, and TSHC06.

\subsection{Created pairs in the inner magnetosphere}
\label{sec:pairs}
Let us devote a little more space to examining 
the particle flux along the open field lines that do {\it not} thread
the gap (i.e., $h_{\rm m}<h<1$).
Since $\Ell$ vanishes,
the created, secondary pairs emit synchrotron photons,
which are capable of cascading into tertiary and higher-generation pairs
by $\gamma$-$\gamma$ or $\gamma$-B collisions.
Examining the cascade, we can calculate the rate of pair creation,
which takes place mainly in the inner magnetosphere. 
Denoting that the pair creation rate is 
$\kappa_{\rm w}(h)\cdot(\Omega B/2\pi e)$ per unit area per second,
we find $\kappa_{\rm w}= 2.2 \times 10^4$,
$2.1 \times 10^4$,
$1.9 \times 10^4$,
$1.8 \times 10^4$,
$1.6 \times 10^4$,
$1.1 \times 10^4$,
$1.0 \times 10^4$,
$0.99\times 10^4$,
$0.95\times 10^4$,
$0.96\times 10^4$,
$1.0 \times 10^4$ at $h=0.20125$, $0.2025$, $0.20375$, $0.205$,
$0.305$, $0.405$, $0.505$, $0.605$, $0.705$, $0.805$, and $0.905$,
respectively, for $\inc=70^\circ$, 
$\mu= 6.0\times 10^{30}\mbox{ G cm}^3$, and $h_{\rm m}=0.200$.
Thus, the averaged creation rate becomes 
${\bar\kappa}_{\rm w}= 1.4 \times 10^4(\Omega B/2\pi e)$ 
pairs per unit area per second,
giving $\dot{N}_{\rm pair}= 3.8 \times 10^{38} \mbox{s}^{-1}$
as the pair creation rate in the entire magnetosphere.
It should be noted that this $\dot{N}_{\rm pair}$ appears less than 
the constraints that arise from consideration of
magnetic dissipation in the wind zone
(Kirk \& Lyubarsky~2001, who derived $10^{40}\,\mbox{s}^{-1}$),
and of Crab Nebula's radio synchrotron emission
(Arons 2004, who derived $10^{40.5}\,\mbox{s}^{-1}$).

Due to strong synchrotron radiation, 
these inwardly created particles quickly lose energy 
to fall onto the lowest Landau level,
preserving longitudinal momentum $\sim m_{\rm e}c$ per particle.
These non-relativistic particles possess momentum flux of
$ 2\dot{N}_{\rm pair} m_{\rm e}c/(\pi R_{\rm PC}^2)
  \sim 1.0\times 10^{12} \mbox{dyn cm}^{-2}$,
where $R_{\rm PC}\sim \sqrt{r_\ast^3/\rlc}$ denotes the polar cap radius.
On the other hand, surface X-ray field has the upward momentum flux of
$6.0 \times 10^8 (kT/100\mbox{eV})^4 \mbox{dyn cm}^{-2}$.
Thus, the created pairs will not be pushed back by
resonant scatterings.
They simply fall onto the stellar surface
with non-relativistic velocities.
The luminosity of $e^-$-$e^+$ annihilation line is about
$3\times10^{26}\mbox{\,ergs s}^{-1}$,
which is negligible 
(e.g., compared with the $\gamma$-ray luminosity 
 $\sim 10^{34.5}\mbox{\,ergs s}^{-1}$).
On these grounds, for the Crab-pulsar,
we must conclude that the present work 
fails to explain the injection rate of the wind particles,
in the same way as in other outer-gap models.

\subsection{Comparison with Polar-slot gap model }
\label{sec:SG}
It is worth comparing the present results with the polar-slot-gap model
proposed by Muslimov and Harding (2003; MH04a; MH04b),
who obtained a quite different solution 
(e.g., negative $\Ell$ in the gap)
solving essentially the same equations 
under analogous boundary conditions
for the same pulsar as in the present work.
The only difference is the transfield thickness of the gap.
Estimating the transfield thickness to be
$\Delta l_{\rm SG} \sim h_{\rm m} r_\ast \sqrt{r/\rlc}$,
which is a few hundred times thinner than the present work,
they extended the solution (near the polar cap surface) 
that was obtained by MT92
into the higher altitudes (towards the light cylinder).
Because of this very small $\Delta l_{\rm SG}$,
emitted $\gamma$-rays do not efficiently materialize within the gap;
as a result, the created and returned positrons from the higher altitudes
do not break down the original assumption of 
the completely-charge-separated SCLF near the stellar surface.

To avoid the reversal of $\Ell$ in the gap
(from negative near the star to positive in the outer magnetosphere),
or equivalently,
to avoid the reversal of the sign of the effective charge density,
$\rho_{\rm eff} = \rho-\rhoGJ$, along the field line,
MH04a and MH04b assumed that
$\rho_{\rm eff}/B$ nearly vanishes and
remains constant above a certain altitude, $s=s_{\rm c}$,
where $s_{\rm c}$ is estimated to be within a few neutron star radii.
Because of this assumption, $\Ell$ is suppress at a very small value
and the pair creation becomes negligible in the entire gap. 
In another word, the enhanced screening is caused not only
by the proximity of two conducting boundaries,
but also by the assumption of 
$\partial(\rho_{\rm eff}/B)/\partial s=0$ within the gap
(see eq.~[\ref{eq:CHR}]).
To justify this $\rho/B$ distribution,
MH04a and MH04b proposed an idea that $\rho$ should grow by 
the {\it cross field motion} of charges due to the toroidal forces,
and that $\rho_{\rm eff}/B$ is a small constant so that 
$c\rho_{\rm eff}/B$ may not exceed the flux of the 
emitted charges from the star,
which ensures the {\it equipotentiality} of the slot-gap boundaries
(see \S~2.2 of MH04a for details).

The cross-field motion becomes important
if particles gain angular momenta as they migrate outwards 
to pick up energies 
which is a non-negligible fraction of the difference 
of the cross-field potential between the two conducting boundaries.
Denoting the fraction as $\epsilon$, we obtain
$\Gamma m_{\rm e}c^2 \dot\varphi \Omega (\varpi/c)^2 
 =\epsilon\,eB \Delta l_{\rm SG}$
(Mestel~1985; eq.~[12] of MH04a).
If we substitute their estimate $\Delta l_{\rm SG} \sim r_\ast/20$,
we obtain 
$\epsilon \sim 0.33 (\dot\varphi/\Omega)\gamma_7 B_{6}^{-1} r_6^{-3}
 (\varpi/\rlc)^2$,
where $\gamma_7=\Gamma/10^7$, $B_{6}=B_\ast/10^{6}\,\mbox{G}$,
and $r_6=r_\ast/10\,\mbox{km}$;
therefore, the cross-field motion becomes important 
in the outer magnetosphere within their slot-gap model.
A larger value of $\Delta l_{\rm SG}$ is incompatible
with the constancy of $\rho_{\rm eff}/B$ due to the cross-field motion
in the higher altitudes.

As for the equipotentiality of the boundaries,
it seems reasonable to suppose that 
$c\vert\rho_{\rm eff}\vert/B < c\vert\rho_\ast\vert/B_\ast$ 
should be held at any altitudes in the gap, as MH04a suggested,
where $\rho_\ast$ denotes the real charge density 
in the vicinity of the stellar surface.
However, the assumption that $\rho_{\rm eff}/B$ is a small positive
constant may be too strong, because it is only a sufficient condition
of $c\vert\rho_{\rm eff}\vert/B < c\vert\rho_\ast\vert/B_\ast$.

In the present paper, on the other hand,
we assume that the magnetic fluxes threading the gap is unchanged,
considering that charges freely move along the field lines on the
upper (and lower) boundaries.
As a result, the gap becomes much thicker than MH04a,b; namely,
$\Delta l_{\rm SG} \sim 0.5 h_{\rm m} \rlc$,
which gives $\epsilon < 10^{-3}$.
Therefore, we can neglect the cross-field motion and justify
the constancy of $\rho/B$ in the outer region of the gap,
where pair creation is negligible.
In the inner magnetosphere, 
$\rho_{\rm eff}/B$ becomes approximately a negative constant,
owing to the discharge of the copiously created pairs.
Because of this negativity of $\rho_{\rm eff}/B$, 
a positive $\Ell$ is exerted.
For a super-GJ solution, we obtain
$j_{\rm e}+j_{\rm ion} \sim 0.9 > \rho_{\rm eff}/(\Omega B/2\pi)$,
which guarantees the equipotentiality of the boundaries.
For a sub-GJ solution, a problem may occur regarding the equipotentiality;
nevertheless, we are not interested in this kind of solutions.

In short, whether the gap solution becomes 
MH04a way 
(with a negative $\Ell$ as an outward extension of the polar-cap model)
or this-work way 
(with a positive $\Ell$ as an inward extension of the outer-gap model)
entirely depends on the transfield thickness and on 
the resultant $\rho_{\rm eff}/B$ variation along the field lines.
If $\Delta l_{\rm SG} \sim r_\ast/10$ holds in the outer magnetosphere,
$\rho_{\rm eff}/B$ could be a small positive constant 
by the cross-field motion of charges (without pair creation);
in this case, the current is slightly sub-GJ with electron emission 
from the neutron star surface,
as MH04a,b suggested.
On the other hand, if $\Delta l_{\rm SG} > \rlc/40$ holds
in the outer magnetosphere,
$\rho_{\rm eff}/B$ takes a small negative value
by the discharge of the created pairs
(see fig.~\ref{fig:C70nine});
in this case, the current is super-GJ with ion emission from the surface,
as demonstrated in the present paper.
Since no studies have ever successfully constrained 
the gap transfield thickness,
there is room for further investigation on this issue.

\acknowledgments
The author is grateful to Drs.
J.~G.~Kirk, B.~Rudak, S.~Shibata, K.~S.~Cheng, 
A.~K.~Harding, J.~Arons, R.~Taam, H.~K.~Chang, and J.~Takata 
for helpful suggestions.
Some important part of this work was prepared while the author studied
at Max-Planck-Institut f\"ur Kernphysik, Heidelberg.
This work is supported by the Theoretical Institute for
Advanced Research in Astrophysics (TIARA) operated under Academia Sinica
and the National Science Council Excellence Projects program in Taiwan
administered through grant number NSC 94-2752-M-007-001.
Also, this work is partly supported by
KBN through the grant 2P03D.004.24 to B.~Rudak,
which enabled the author to use the MEDUSA cluster at CAMK Toru\'n.
The final computations were carried out
with the aid of the Blade Tank servers at TIARA, Taipei.

\acknowledgments

\appendix

\section{Appendix}
\label{sec:app}
Explicit expressions of equations~(\ref{eq:def_mag1})
and (\ref{eq:def_mag2}) are as follows:

\begin{equation}
  g^{ss}
  = g^{rr}
    \left(\frac{\partial s}{\partial r}\right)_{\theta,\varphi}^2
   +g^{\theta\theta}
    \left(\frac{\partial s}{\partial \theta}\right)_{\varphi,r}^2
   -\frac{k_0}{\rhowSQR}
    \left(\frac{\partial s}{\partial \varphi}\right)_{r,\theta}^2
\end{equation}
\begin{equation}
  g^{\theta_\ast \theta_\ast}
  = g^{rr}
    \left(\frac{\partial \theta_\ast}
               {\partial r}\right)_{\theta,\varphi}^2
   +g^{\theta\theta}
    \left(\frac{\partial \theta_\ast}
               {\partial \theta}\right)_{\varphi,r}^2
   -\frac{k_0}{\rhowSQR}
    \left(\frac{\partial \theta_\ast}
               {\partial \varphi}\right)_{r,\theta}^2
\end{equation}
\begin{equation}
  g^{\varphi_\ast \varphi_\ast}
  = g^{rr}
    \left(\frac{\partial \varphi_\ast}
               {\partial r}\right)_{\theta,\varphi}^2
   +g^{\theta\theta}
    \left(\frac{\partial \varphi_\ast}
               {\partial \theta}\right)_{\varphi,r}^2
   -\frac{k_0}{\rhowSQR}
    \left(\frac{\partial \varphi_\ast}
               {\partial \varphi}\right)_{r,\theta}^2
\end{equation}
\begin{equation}
  g^{s\theta_\ast}
  = g^{rr}
    \left(\frac{\partial s}
               {\partial r}\right)_{\theta,\varphi}
    \left(\frac{\partial \theta_\ast}
               {\partial r}\right)_{\theta,\varphi}
   +g^{\theta\theta}
    \left(\frac{\partial s}
               {\partial \theta}\right)_{\varphi,r}
    \left(\frac{\partial \theta_\ast}
               {\partial \theta}\right)_{\varphi,r}
   -\frac{k_0}{\rhowSQR}
    \left(\frac{\partial s}
               {\partial \varphi}\right)_{r,\theta}
    \left(\frac{\partial \theta_\ast}
               {\partial \varphi}\right)_{r,\theta},
\end{equation}
\begin{equation}
  g^{\theta_\ast \varphi_\ast}
  = g^{rr}
    \left(\frac{\partial \theta_\ast}
               {\partial r}\right)_{\theta,\varphi}
    \left(\frac{\partial \varphi_\ast}
               {\partial r}\right)_{\theta,\varphi}
   +g^{\theta\theta}
    \left(\frac{\partial \theta_\ast}
               {\partial \theta}\right)_{\varphi,r}
    \left(\frac{\partial \varphi_\ast}
               {\partial \theta}\right)_{\varphi,r}
   -\frac{k_0}{\rhowSQR}
    \left(\frac{\partial \theta_\ast}
               {\partial \varphi}\right)_{r,\theta}
    \left(\frac{\partial \varphi_\ast}
               {\partial \varphi}\right)_{r,\theta},
\end{equation}
\begin{equation}
  g^{\varphi_\ast s}
  = g^{rr}
    \left(\frac{\partial \varphi_\ast}
               {\partial r}\right)_{\theta,\varphi}
    \left(\frac{\partial s}
               {\partial r}\right)_{\theta,\varphi}
   +g^{\theta\theta}
    \left(\frac{\partial \varphi_\ast}
               {\partial \theta}\right)_{\varphi,r}
    \left(\frac{\partial s}
               {\partial \theta}\right)_{\varphi,r}
   -\frac{k_0}{\rhowSQR}
    \left(\frac{\partial \varphi_\ast}
               {\partial \varphi}\right)_{r,\theta}
    \left(\frac{\partial s}
               {\partial \varphi}\right)_{r,\theta},
\end{equation}
and 
\begin{equation}
  A^s \equiv
  \frac{c^2}{\sqrt{-g}}
   \left\{ \partial_r \left[ \frac{g_{\varphi\varphi}}{\rhowSQR}
                             \sqrt{-g}g^{rr}
                             \left(\frac{\partial s}
                                        {\partial r}\right)_{\theta,\varphi}
                      \right]
          +\partial_\theta
                      \left[ \frac{g_{\varphi\varphi}}{\rhowSQR}
                             \sqrt{-g}g^{\theta\theta}
                             \left(\frac{\partial s}
                                        {\partial \theta}\right)_{\varphi,r}
                      \right]
    \right\}
   -\frac{c^2 g_{\varphi\varphi}}{\rhowSQR}
    \frac{k_0}{\rhowSQR}
    \left(\frac{\partial^2 s}{\partial\varphi^2}\right)_{r,\theta},
\end{equation}
\begin{equation}
  A^{\theta_\ast} \equiv
  \frac{c^2}{\sqrt{-g}}
   \left\{ \partial_r \left[ \frac{g_{\varphi\varphi}}{\rhowSQR}
                             \sqrt{-g}g^{rr}
                             \left(\frac{\partial \theta_\ast}
                                        {\partial r}\right)_{\theta,\varphi}
                      \right]
          +\partial_\theta
                      \left[ \frac{g_{\varphi\varphi}}{\rhowSQR}
                             \sqrt{-g}g^{\theta\theta}
                             \left(\frac{\partial \theta_\ast}
                                        {\partial \theta}\right)_{\varphi,r}
                      \right]
    \right\}
   -\frac{c^2 g_{\varphi\varphi}}{\rhowSQR}
    \frac{k_0}{\rhowSQR}
    \left(\frac{\partial^2 \theta_\ast}
               {\partial\varphi^2}\right)_{r,\theta},
\end{equation}
\begin{equation}
  A^{\varphi_\ast} \equiv
  \frac{c^2}{\sqrt{-g}}
   \left\{ \partial_r \left[ \frac{g_{\varphi\varphi}}{\rhowSQR}
                             \sqrt{-g}g^{rr}
                             \left(\frac{\partial \varphi_\ast}
                                        {\partial r}\right)_{\theta,\varphi}
                      \right]
          +\partial_\theta
                      \left[ \frac{g_{\varphi\varphi}}{\rhowSQR}
                             \sqrt{-g}g^{\theta\theta}
                             \left(\frac{\partial \varphi_\ast}
                                        {\partial \theta}\right)_{\varphi,r}
                      \right]
    \right\}
   -\frac{c^2 g_{\varphi\varphi}}{\rhowSQR}
    \frac{k_0}{\rhowSQR}
    \left(\frac{\partial^2 \varphi_\ast}
               {\partial\varphi^2}\right)_{r,\theta}.
\end{equation}


\begin{thebibliography}{}
\bibitem[Arons(1983)]{aron83} 
    Arons, J. 1983, ApJ 302, 301
\bibitem[Arons \& Scharlemann(1979)]{aron79} 
    Arons, J., Scharlemann, E. T. 1979, ApJ 231, 854
\bibitem{Arons04}
    J. Arons 2004, Adv. in Sp. Res. 33, 466
\bibitem[Becker \& Tr\"umper(1997)]{beck97} 
    Becker, W., Tr\"umper, J. 1997, A\&A 326, 682 
\bibitem[Beskin et al.(1992)]{besk92} 
    Beskin, V. S., Istomin, Ya. N., \& Par'ev, V. I. 
    1992, Sov. Astron. 36(6), 642
\bibitem[Camenzind(1986a)]{came86a} 
  Camenzind, M. A \& A, 156, 137
\bibitem[Camenzind(1986b)]{came86b} 
  Camenzind, M. A \& A, 162, 32
\bibitem[Cheng et al.(1986a)]{chen86a} 
    Cheng, K. S., Ho, C., \& Ruderman, M., 1986a
    \apj, 300, 500 (CHR86a)
\bibitem[Cheng et al.(1986b)]{chen86b} 
    Cheng, K. S., Ho, C., \& Ruderman, M., 1986b
    \apj, 300, 522 (CHR86b)
\bibitem[Cheng \& Zhang(1996)]{chen96} 
    Cheng, K. S., \& Zhang, L. 1996, \apj 463, 271
\bibitem[Cheng et al.(2000)]{chen00} 
    Cheng, K. S., Ruderman, M., \& Zhang, L. 2000,
    \apj, 537, 964
\bibitem[Chiang \& Romani(1992)]{chia92} 
    Chiang, J., \&  Romani, R. W. 1992,
    \apj, 400, 629
\bibitem[Chiang \& Romani(1994)]{chia94} 
    Chiang, J., \& Romani, R. W. 1994,
    \apj, 436, 754
\bibitem[Contopoulos et al(1999)]{cont99} 
   Contopoulos, I., Kazanas, D., and Fendt, C. 1999,
   \apj, 511, 351
\bibitem[Daugherty \& Harding(1982)]{Daug82} 
    Daugherty, J. K., \& Harding, A. K. 1982, 
    \apj, 252, 337
\bibitem[Daugherty \& Harding(1996)]{Daug96} 
    Daugherty, J. K., \& Harding, A. K. 1996, 
    \apj, 458, 278
\bibitem[de Naurois et al.(2002)]{naur02} 
    de~Naurois, M., Holder, J., Bazer-Bachi, R., Bergeret, H.,
    Bruel, P., Cordier, A., Debiais, G., Dezalay, J.-P., et al. 2002,
    \apj 566, 343
\bibitem[Dermer(1994)]{derm94} 
    Dermer, C. D., \& Sturner, S. J. 1994, 
    \apjl, 420, L75
\bibitem[Erber(1966)]{erber66} 
    Erber, T. 1966, Rev. Mod. Phys., 38, 626
\bibitem[Diks \& Rudak(2003)]{dyks03} 
    Dyks, J., \& Rudak, B. 2003, \apj 598, 1201
\bibitem[Diks et al.(2004)]{dyks04} 
    Dyks, J., Harding, A. K., \& Rudak, B. 2004, \apj 606, 1125
\bibitem[Fierro et al.(1998)]{fier98} 
    Fierro, J. M., Michelson, P. F., Nolan, P. L., \& Thompson, D. J.,
    1998, \apj 494, 734
\bibitem[Goldreich \& Julian(1969)]{gold69} 
    Goldreich, P. Julian, W. H.
    1969, \apj. 157, 869
\bibitem{Goodwin04} 
    Goodwin, S. P. Mestel, J. Mestel, L. Wright G. A. E. 
    2004, MNRAS 349, 213
\bibitem[Gruzinov(2005)]{gruz05} 
    Gruzinov, A. 2005, Phys. Rev. Letters 94, 021101
\bibitem[Harding et al.(1978)]{hard78} 
    Harding, A. K., Tademaru, E., \& Esposito, L. S. 1978, 
    \apj, 225, 226
 \bibitem[Hirotani(2001)]{hiro01} 
     Hirotani, K. 2001, \apj 549, 495 
 \bibitem[Hirotani(2006)]{hiro06} 
     Hirotani, K. 2006, Mod. Phys. Lett. A (Brief Review) 21, 1319--1337
     (astroph/0606017).
\bibitem[Hirotani \& Shibata(1999a)]{hiro99a} 
    Hirotani, K. \& Shibata, S.,
    1999a, \mnras 308, 54 
\bibitem[Hirotani \& Shibata(1999b)]{hiro99b} 
    Hirotani, K. \& Shibata, S.,
    1999b, \mnras 308, 67 
\bibitem[Hirotani \& Shibata(1999c)]{hiro99c} 
    Hirotani, K. \&  Shibata, S.,
    1999c, PASJ 51, 683   
\bibitem[Hirotani \& Shibata(2001a)]{hiro01a} 
    Hirotani, K. \& Shibata, S.,
    2001a, \mnras 325, 1228 
\bibitem[Hirotani \& Shibata(2001b)]{hiro01b} 
    Hirotani, K. \& Shibata, S.,
    2001b, \apj 558, 216 (Paper~VIII)
\bibitem[Hirotani \& Shibata(2002)]{hiro02} 
    Hirotani, K. \& Shibata, S.,
    2002, \apj 564, 369 (Paper~IX)
\bibitem[Hirotani et al.(2003)]{hiro03}
    Hirotani, K., Harding, A. K., \& Shibata, S.,
    2003, \apj 591, 334 (HHS03)
\bibitem[Jackson (1962)]{jack62}
    Jackson, J. D. 1962, Classical electrodynamics
    (New York: John Wiley \& Sons), 591
\bibitem[Jones(1985)]{jones85} 
  Jones, P. B. 1985, Phys. Rev. Lett., 55, 1338
\bibitem[Kanbach (1999)]{kanb99} 
    Kanbach, G. 1999,
    in proc. of the Third INTEGRAL Workshop,
    ed. Bazzaro, A., Palumbo, G. G. C., \& Winkler, C.
    Astrophys. Lett. Comm. 38, 17
\bibitem[Kirk \& Skjaeraasen (2003)]{kirk03} 
    Kirk, J., Skj{\ae}raasen, O. 2003,
    \apj 591, 366
\bibitem[Kuiper et al.(2001)]{kuip01} 
    Kuiper, L., Hermsen, W., Cusumano, G., Riehl, R.
    Sch\"onfelder, V., Strong, A., Bennett, K., \& McConnell, M. L.
    2001, A\& A 378, 918 
\bibitem[Lense and Thirring(1918)]{len18}
    Lense, J. and Thirring, H. 1918,
    Phys. Z. 19, 156.
  Translated by B. Mashhoon F.W. Hehl and D.S. Theiss (1984),
  Gen. Relativ. Gravit. 16, 711.
\bibitem[Mestel(1971)]{mest71} 
    Mestel, L. 1971, Nature Phys. Sci., 233, 149
\bibitem[Mestel et al.(1985)]{mest85} 
    Mestel, L., Robertson, J. A., Wang, Y. M., Westfold, K. C.
    1985, \mnras 217, 443
\bibitem[Morini(1983)]{mori83} 
    Morini, M. 1983, MNRAS, 202, 495
\bibitem[Muslimov \& Tsygan(1992)]{musl92} 
    Muslimov, A. G., \& Tsygan, A. I.
    \mnras, 255, 61 (MT92)
\bibitem[Muslimov \& Harding(2003)]{musl03} 
    Muslimov, A. G., \& Harding, A. K., 2003,
    \apj, 588, 430
\bibitem[Muslimov \& Harding(2004)]{musl04a} 
    Muslimov, A. G., \& Harding, A. K., 2004a,
    \apj, 606, 1143 (MH04a)
\bibitem[Muslimov \& Harding(2004)]{musl04b} 
    Muslimov, A. G., \& Harding, A. K., 2004b,
    \apj, 617, 471 (MH04b)
\bibitem[Muslimov \& Harding(2005)]{musl05} 
    Muslimov, A. G., \& Harding, A. K., 2005,
    \apj, 630, 454
\bibitem[Nakamura(1999)]{naka99} 
    Nakamura, T., Yabe, T., Comput. Phys. Comm. 120, 122 
\bibitem[Neuhauser(1986)]{neuh86} 
    Neuhauser, D., Langanke, K., Koonin, S. E., 1986, 
    Phys. Rev. A33, 2084
\bibitem[Neuhauser(1987)]{neuh87} 
    Neuhauser, D., Koonin, S. E., Langanke, K., 1986, 
    Phys. Rev. A36, 4163
\bibitem[Nolan(1993)]{nola93} 
    Nolan, P. L., Arzoumanian, Z., Bertsch, D. L., 
    Chiang, J., Fichtel, C. E., Fierro, J. M., 
    Hartman, R. C., Hunter, S. D., et al. 1993, \apj 409, 697
\bibitem[Pandharipande]{Pand75} 
    Pandharipande, V. R., Smith, R. A. 1975, Nucl. Phys. A175, 225
\bibitem[Romani(1996)]{roma96} 
    Romani, R. W. 1996, \apj, 470, 469
\bibitem[Romani \& Yadigaroglu(1995)]{roma95} 
    Romani, R. W., \& Yadigaroglu, I. A. 1995,
    \apj 438, 314
\bibitem[Scharlemann et al.(1978)]{shar78} 
    Scharlemann, E. T., Arons, J., \& Fawley, W. T., 1978
    \apj, 222, 297 (SAF78)
\bibitem[Serot(1979a)]{serot79a} 
    Serot, B. D. 1979a, Phys. Letters, 86B, 146
\bibitem[Serot(1979b)]{serot79b} 
    Serot, B. D. 1979b, Phys. Letters, 87B, 403
\bibitem[Shibata(1997)]{shib97} 
    Shibata, S. 1997, \mnras 287, 262
\bibitem{Spit06} 
    A. Spitkovsky 2006, ApJ in press
\bibitem[Sturner et al.(1995)]{stur95} 
    Sturner, S. J., Dermer, C. D., \& Michel, F. C. 1995, 
    \apj 445, 736
\bibitem[Takahashi et al.(1990)]{taka90} 
  Takahashi, M., Nitta, S., Tatematsu, Y., Tomimatsu, A.
  \apj 363, 206 
\bibitem[Takata et al.(2004)]{taka04b} 
    Takata, J., Shibata, S., \& Hirotani, K., 2004,
    \mnras 354, 1120 (TSH04)
\bibitem[Takata et al.(2006)]{taka06} 
    Takata, J., Shibata, S., Hirotani, K., \& Chang, H.-K., 2006,
    \mnras 366, 1310 (TSHC06)
\bibitem[Tennant et al.(2001)]{tenn01} 
    Tennant, A. F., Becker, W., Juda, M., Elsner, R. F.,
    Kolodziejczak, J. J., Murray, S. S., O'Dell, S. L.,
    Paerels, F., Swartz, D. A., Shibazaki, N., Weisskopf, M. C., 2001,
    \apjl 554, 173
\bibitem[Thompson et al.(2001)]{thom01} 
  Thompson, D. J. 2001, in AIP Conf. Proc. 558,
  High Energy Gamma-Ray Astronomy, 
  ed. A. Goldwurm et al. (New York: AIP), 103
\bibitem[Ulmer et al.(1995)]{ulmer95}
  Ulmer, M.~P., Matz, S.~M., Grabelsky, D.~A., Grove, J.~E.,
  Stirickman, M.~S., Much, R., Besetta, M.~C., Strong, A. et al.
  1995, \apj 448, 356 ! --364
\bibitem[Weber \& Davis(1967)]{webe67}
  Weber, E. J., Davis, Leverrett, Jr. \apj 148, 217 
\bibitem[Zhang \& Cheng(1997)]{Lzhan97} 
    Zhang, L. Cheng, K. S. 1997, \apj 487, 370
\bibitem[Znajek(1977)]{Znaj77} 
  Znajek, R. L. \mnras 179, 457
\end{thebibliography}
\end{document}